\documentclass[twocolumn]{aastex631}

\usepackage{amsmath}

\received{October 11, 2024}
\revised{March 17, 2024}
\accepted{June 1, 2025}

\shorttitle{Velocities for a Trail without Dark Matter}
\shortauthors{Keim et al.}

\begin{document}

\title{Kinematic Confirmation of a Remarkable Linear Trail of Galaxies in the NGC 1052 Field, Consistent with Formation in a High-Speed Bullet Dwarf Collision}

\correspondingauthor{Michael A. Keim}
\email{michael.keim@yale.edu}

\author[0000-0002-7743-2501]{Michael A. Keim}
\affiliation{Department of Astronomy, Yale University, PO Box 208101, New Haven, CT 06520-8101, USA}

\author[0000-0002-8282-9888]{Pieter van Dokkum}
\affiliation{Department of Astronomy, Yale University, PO Box 208101, New Haven, CT 06520-8101, USA}

\author[0000-0002-5120-1684]{Zili Shen}
\affiliation{Department of Astronomy, Yale University, PO Box 208101, New Haven, CT 06520-8101, USA}

\author[0000-0001-5079-1865]{Harrison Souchereau}
\affiliation{Department of Astronomy, Yale University, PO Box 208101, New Haven, CT 06520-8101, USA}

\author[0000-0002-7075-9931]{Imad Pasha}
\affiliation{Department of Astronomy, Yale University, PO Box 208101, New Haven, CT 06520-8101, USA}

\author[0000-0002-1841-2252]{Shany Danieli}
\affiliation{Department of Astrophysical Sciences, 4 Ivy Lane, Princeton University, Princeton, NJ 08544}

\author[0000-0002-4542-921X]{Roberto Abraham}
\affiliation{Department of Astronomy \& Astrophysics, University of Toronto, 50 St. George St., Toronto, ON M5S 3H4, Canada}
\affiliation{Dunlap Institute for Astronomy and Astrophysics, University of Toronto, Toronto ON, M5S 3H4, Canada}

\author[0000-0003-2473-0369]{Aaron J. Romanowsky}
\affiliation{Department of Physics \& Astronomy, San Jose State University, One Washington Square, San Jos\'e, CA 95192, USA}
\affiliation{Department of Astronomy \& Astrophysics, University of California Santa Cruz, 1156 High Street, Santa Cruz, CA 95064, USA}

\author[0000-0003-2876-577X]{Yimeng Tang}
\affiliation{Department of Astronomy \& Astrophysics, University of California Santa Cruz, 1156 High Street, Santa Cruz, CA 95064, USA}
 
\begin{abstract}
A unique linear trail of diffuse galaxies was recently identified in the NGC 1052 field. This trail includes the remarkable, ultra-diffuse galaxies DF2 and DF4 which lack dark matter and host unusually luminous globular clusters. It has been proposed that the trail formed via a high-speed collision between two gas-rich dwarf galaxies. This scenario predicts that the trail galaxies are kinematically connected and follow a specific trend in radial velocity as a function of position, based on the known velocities and positions of DF2 and DF4. To test this hypothesis, we measured radial velocities for seven additional galaxies on the trail. While the galaxies' low surface brightnesses presented observational challenges, we employ several methods to obtain measurements for galaxies with effective surface brightnesses up to 28.6 mag arcsec$^{-2}$, including a narrow slit placed over globular clusters and a novel wide slit mode on Keck/LRIS, as well as a `light bucket' mode on Keck/KCWI. We find that five of our seven targets follow the precise velocity trend predicted by DF2 and DF4, to a degree with just a 2\% chance of randomly occurring. Moreover, the trail galaxies' radial velocities are significantly higher than those of the NGC 1052 group, setting it apart as a separate, kinematically connected system. Our findings support the theory that this trail of galaxies, including DF2 and DF4, formed together in a single event. A `bullet dwarf' collision remains the only known explanation for all the unusual properties of DF2, DF4, and the associated trail of galaxies.
\end{abstract} 



\section{Introduction} \label{Sec:Introduction}

Two galaxies in the NGC 1052 field, NGC 1052-DF2 and NGC 1052-DF4 (DF2 and DF4, hereafter), share several exceptional properties that, taken together, set them apart from all other galaxies previously known. First, they host populations of extremely luminous globular clusters $\approx$1.5 magnitudes brighter than `universal' values \citep{2018ApJ...856L..30V,2019ApJ...874L...5V,2021ApJ...909..179S}. Second, they are large with low central surface brightnesses, falling into the category of ultra-diffuse galaxies \citep{2015ApJ...798L..45V}. Finally, they both have little to no dark matter, with dynamical masses that are consistent with their stellar masses alone \citep{2018Natur.555..629V,2018ApJ...863L..15W,2019ApJ...874L...5V,2019ApJ...874L..12D,2019A&A...625A..76E,2019ApJ...877..133D,2022ApJ...935..160K,2023ApJ...957....6S}.

\citet{2019MNRAS.488L..24S} proposed that DF2 formed as the result of a high velocity collision between two gas-rich dwarf progenitors. In a dwarf galaxy scale analog of the Bullet Cluster \citep{2006ApJ...648L.109C}, this `bullet dwarf' event would cause the collisionless stars and dark matter of the progenitor galaxies to separate out from their gas, which may go on to form dark matter free galaxies. Simulations show that these unusual conditions favor the formation of massive globular clusters like those seen in DF2 and DF4 (\citealt{2020ApJ...899...25S,2021ApJ...917L..15L}; similar theory work also suggests that the large sizes of the galaxies can be explained by feedback; \citealt{2022MNRAS.510.3356T}). 

Although such a collision is unlikely to occur twice in the same group, DF2 and DF4 have relative radial velocities, relative distances, and ages consistent with the joint formation of both galaxies in a single high-speed collision \citep{2022Natur.605..435V}. What's more, \citet{2022Natur.605..435V} identified novel evidence which suggests DF2 and DF4 formed together in a collision: there is a tight, highly statistically significant linear alignment of a dozen faint, spatially extended objects near NGC 1052, including DF2 and DF4. Such an alignment of galaxies was evident in the results of prior simulations, which predicted that gas separated out by a bullet dwarf collision would fragment into a linear trail of several additional small dark matter free galaxies, as well as DF2 and DF4 like objects (\citealt{2020ApJ...899...25S,2021ApJ...917L..15L}). More recent simulations with realistic gravitohydrodynamic conditions and orbital trajectories \citep{2024ApJ...966...72L} demonstrate how a bullet collision can indeed reproduce the trail we see today, with a line of dark matter free galaxies with 9$\pm$2 Gyr ages (\citealt{2018ApJ...856L..30V}; \citealt{2019A&A...625A..77F}; \citealt{2025ApJ...978...21T}) consistent with having formed shortly after said collision, stretching over a physical length comparable to the 1.7$\pm$0.5 Mpc distance between DF2 and DF4 \citep{2023ApJ...957....6S}.

In order to further explore this bullet dwarf scenario, \citet{2022ApJ...940L...9V} conducted a basic hypothesis-falsification test by examining the colors of globular clusters in DF2 and DF4. Given that the globular clusters formed from similar collided, shock-compressed gas and were both theorized and shown in simulation to form rapidly with high efficiency \citep{2019MNRAS.488L..24S,2021ApJ...917L..15L}, the globular clusters ought to have similar ages, metallicities, and therefore, colors. Indeed, careful analysis of deep \textit{Hubble Space Telescope (HST)} imaging showed that the globular clusters were uniquely monochromatic compared to general dwarf populations, both between the two galaxies and within the galaxies themselves \citep{2022ApJ...940L...9V}.

Another important prediction of this bullet dwarf collision scenario is that the radial velocities of galaxies on the linear trail should follow a specific kinematic trend as a function of position along the line \citep{2022ApJ...940L...9V,2024ApJ...966...72L}, with galaxies closer to DF2 ($cz$ = 1805$^{+1}_{-1}$ km s$^{-1}$; \citealt{2019ApJ...874L..12D}) being at higher radial velocities and galaxies closer to DF4 ($cz$ = 1433.3$^{+0.3}_{-0.4}$ km s$^{-1}$; \citealt{2023ApJ...957....6S}) being at lower radial velocities. Although $2\pm2$ of the trail galaxies are expected to be random interlopers (\citealt{2022Natur.605..435V}; in particular, the brighter, more compact galaxy LEDA 4014647), most of the galaxies on the trail should follow this precise trend. Identification of this velocity trend would mean the trail, currently a unique geometrical arrangement, is also a kinematically connected system, consistent with expectations for their joint, collisional formation.

The bullet dwarf model is of great interest, representing a novel, top-down galaxy formation channel involving intense star formation in systems without dark matter. Observations of bullet dwarf systems are crucial as such extreme environments are not well modeled in large scale cosmological simulation \citep{2020ApJ...899...25S}. Moreover, the bullet dwarf scenario represents a probe into the fundamental nature of dark matter: it relies on dark matter being a largely collisionless particle and, as in the Bullet Cluster, cannot be explained by modifying gravity alone \citep{2018RvMP...90d5002B,2022Symm...14.1331B,2007ApJ...654L..13A}, challenging modified Newtonian dynamics (MOND; \citealt{1983ApJ...270..365M}) on the same galaxy scales where the theory is most relevant. Moreover, it may provide the ingredients for limits on dark matter self-interaction cross section constraints as in \citet{2008ApJ...679.1173R}, on the scales that are relevant for explaining the dark-matter density profiles of low-mass galaxies \citep{2000PhRvL..84.3760S}. Though hundreds of high speed collisions which would produce dark matter free galaxies have been identified in the large scale cosmological simulation TNG100 \citep{2020ApJ...899...25S}, and high speed galaxy collisions have already been observed to produce linear trails of gas (\citealt{2008ApJ...687L..69K}, which under the standard cosmological model would be assumed to lack dark matter), systems containing dark matter free galaxies produced in this manner have yet to be observed in nature. Examination of the kinematic prediction of the bullet dwarf scenario is, therefore, critical.

Thus, the goal of this paper is to obtain radial velocity measurements of the trail galaxies in order to test the hypothesis that they are kinematically connected and follow the velocity trend predicted by the bullet dwarf collision scenario. Our outline is as follows: in Section~\ref{Sec:Data}, we review our observations and the reduction procedures used in a variety of observing modes.  In Section~\ref{Sec:Results} we report our radial velocity measurements and conduct an analysis of their probabilistic significance. We discuss and contextualize our work in Section~\ref{Sec:Discussion}, and give a final summary in Section~\ref{Sec:Summary and Conclusion}. A toy model simulation reviewing kinematic trend expectations is also given for reference in Appendix~\ref{Sec:ToyModel}.


\section{Data} \label{Sec:Data}

\begin{figure*}
    \centering
    \includegraphics[width=\textwidth]{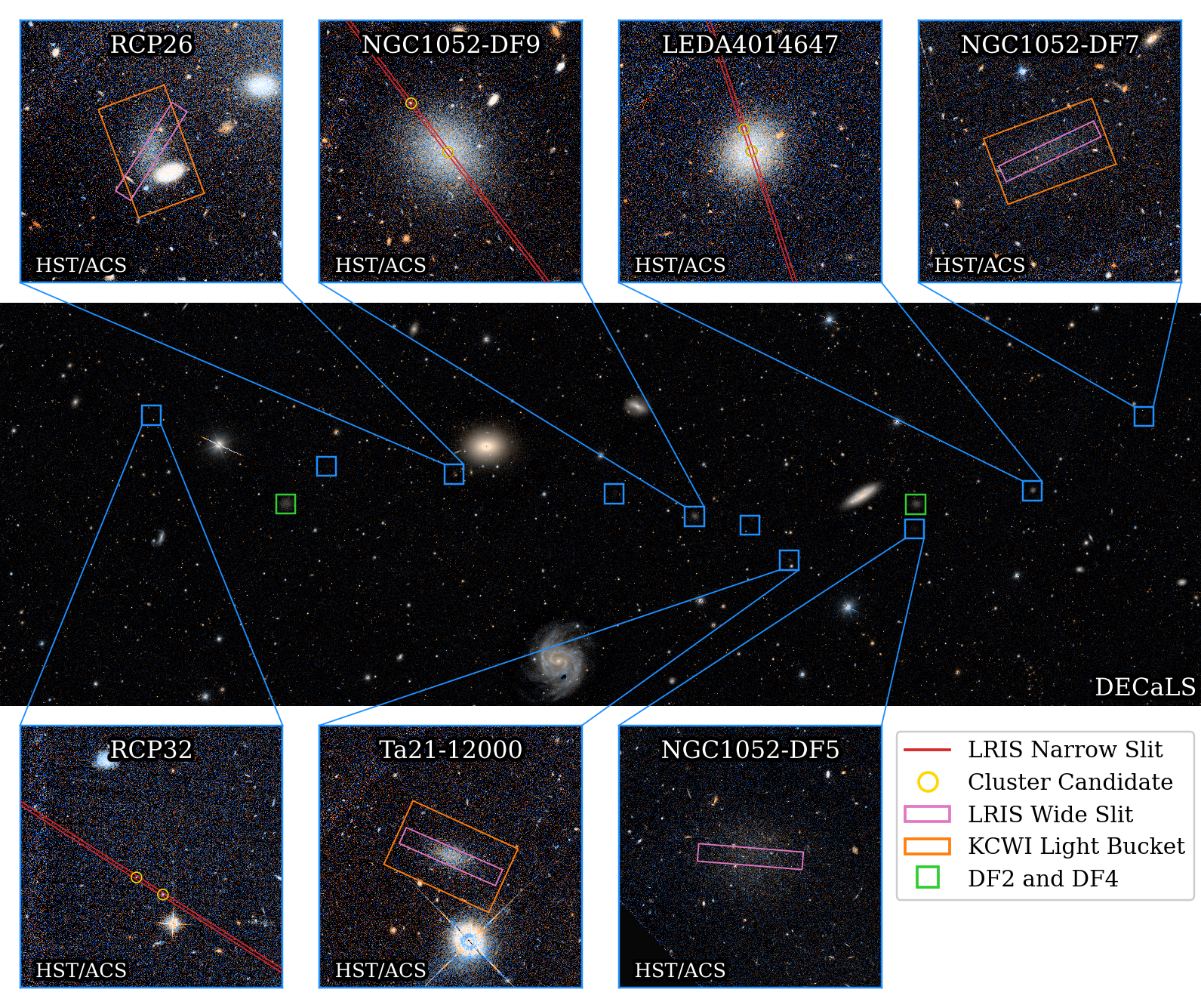}
    \caption{A graphical overview of our observing strategy. Targeted galaxies are shown in $75\arcsec{\times}75\arcsec$ \textit{HST} cutouts, with rectangles indicating the specified field of view for utilized observation modes (\textit{zoom-in panels}). This includes the LRIS narrow slit mode (\textit{red, with targeted globular cluster candidates indicated with yellow circles}), LRIS wide slit mode (\textit{pink}), and KCWI light bucket mode (\textit{orange}). Trail galaxies are indicated in $75\arcsec{\times}75\arcsec$ boxes (\textit{blue, with DF2 and DF4 highlighted in green}) on a DECaLS image of the field. All images are oriented along the line connecting DF2 and DF4.\label{Fig:Observations}}
\end{figure*}

As first identified by \citet{2022Natur.605..435V}, the trail consists of 12 diffuse galaxies mostly discovered in deep surveys searching for low surface brightness objects \citep{2018ApJ...868...96C,2021A&A...656A..44R}. An initial review of the sample, which includes RCP32, DF2, RCP28, RCP26, RCP21, NGC 1052-DF9 (DF9, hereafter), RCP17, Ta21-12000, NGC 1052-DF5 (DF5, hereafter), DF4, LEDA 4014647, and NGC 1052-DF7 (DF7, hereafter), is given in \citet{2022Natur.605..435V}. The galaxies are indicated by blue boxes in the central panel of Figure~\ref{Fig:Observations}. Due to the low surface brightness nature of the galaxies, obtaining radial velocities from redshifts for the galaxies without existing measurements is a non-trivial undertaking, even with large 10 m telescopes at the W. M. Keck Observatory which we will utilize in this work. 

To address this, we first identified globular cluster candidates in \textit{HST} Advanced Camera for Surveys (ACS) imaging from mid-cycle program 16912 (PI: van Dokkum). We then selected two cluster candidates in each of DF9, LEDA 4014647, and RCP32 to target with a `narrow slit' mode observation using the Low Resolution Imaging Spectrometer (LRIS) on Keck I, thereby allowing for relatively fast redshift measurement. The last of these three galaxies stands out as one of the most diffuse galaxies beyond the Local Group ever detected (by \citealt{2021A&A...656A..44R} with an effective surface brightnesses of 28.6 mag arcsec$^{-2}$), making a redshift a highly interesting prospect.

For the remaining diffuse objects, we utilized two other methods to make radial velocity measurements feasible. First, we used a novel `wide slit' observing mode on LRIS, covering the broad Ca II H and K Fraunhofer lines with a high resolution grating and binning heavily. Second, we utilized spectroscopy with the Keck Cosmic Web Imager (KCWI) on Keck II in the `light bucket' mode, combining most of the Integral Field Unit (IFU) into a single spectrum. These methods allowed for redshift measurements of an additional four galaxies in reasonable integration times, including RCP26, Ta21-12000, DF5, and DF7. The remaining three galaxies, RCP17, RCP21, and RCP28, were too faint for redshift measurement even using the above methods. 

A summary of our observing strategy is given in Figure~\ref{Fig:Observations}, with \textit{HST} imaging from program 16912 used to identify globular clusters, with supplementary imaging from programs 14644 and 15695 (PI: van Dokkum), and a cutout from the Dark Energy Camera (DECam) Legacy Survey (DECaLS; \citealt{2019AJ....157..168D}).


\subsection{Observations} \label{Sec:Observation}

As summarized in Table~\ref{Table:Observations}, data for LRIS measurements were taken across 5 nights including 2021 November 2--3, 2022 September 29--30, and 2022 October 1. For narrow slit mode observations of globular cluster candidates we utilized a 1$\arcsec$ longslit with a position angle such that the slit was placed across the two candidates for each galaxy. On the blue side we used the 300 lines mm$^{-1}$ grism blazed at 5000 \r{A} and on the red side we used the 1200 lines mm$^{-1}$ grating blazed at 9000 \r{A}, with the 680 nm dichroic. For wide slit mode observation of other diffuse galaxies we used a wide 5$\arcsec$ slit with the high resolution 1200 lines mm$^{-1}$ grism blazed at 3400 \r{A} on the blue side and the 1200 lines mm$^{-1}$ grating blazed at 7500 \r{A} on the red side, with the 560 nm dichroic. Several observations of DF5 also employed a 7$\arcsec$ slit, with otherwise the same configuration. Conditions were generally good with the exception of poor seeing on the third night during wide slit mode observation of DF7; importantly, seeing during narrow slit mode observation was 0.7--0.8$\arcsec$.

Data for KCWI measurements were taken across 2 nights, 2021 November 9 and 2023 October 3. We utilized the large image slicer with the BL grating at a central wavelength of 4600 \r{A} on the first night and 4500 \r{A} on the second night. As noted in Table~\ref{Table:Observations}, on the second night we included sky offsets before and after every on-target observation. There were intermittent clouds on the first night, seeing was 1.5-1.7$\arcsec$, though this should not strongly affect our science goals with the large slicer observations. Conditions were good on the second night.

\begin{deluxetable}{lcr}
\tablecaption{Observations utilized in this work.\label{Table:Observations}}
\tablewidth{0pt}
\tablehead{
\colhead{Source} & \colhead{Observing Mode}  & \colhead{Exposure Time (s)}
}
\startdata
NGC 1052-DF9   & LRIS Narrow Slit  & 3600\\
LEDA 4014647   & LRIS Narrow Slit  & 7200\\
RCP32         & LRIS Narrow Slit  & 19200\\
NGC 1052-DF5   & LRIS Wide Slit    & 31200\tablenotemark{a}\!\!\!\:\!\!\\
Ta21-12000    & LRIS Wide Slit    & 3600\\
RCP26         & LRIS Wide Slit    & 6000\\
NGC 1052-DF7   & LRIS Wide Slit    & 12000\\
Ta21-12000    & KCWI Light Bucket & 3600\\
RCP26         & KCWI Light Bucket & 1200\\
NGC 1052-DF7   & KCWI Light Bucket & 1800\tablenotemark{b}\!\!\!\:\!\!\\
\enddata
\tablenotetext{a}{This total includes 14400 s taken with a 5$\arcsec$ slit (as used on all other galaxies in the LRIS wide slit observing mode) and 16800 s taken with a 7$\arcsec$ slit.}
\tablenotetext{b}{This total includes only on-target exposure time, whereas for DF7, unlike other targets, we alternated the three 600 s on-target exposures with four 600 s off-target sky exposures, for a total of 4200 s.}
\end{deluxetable}


\subsection{Spectral Reduction} \label{Sec:Reduction}

Extracting 1D spectra from our various spectroscopic observations required several different reduction procedures. For LRIS narrow slit mode observations, we used a custom pipeline, with {\tt PypeIt} \citep{2020JOSS....5.2308P} used for basic sky model, cosmic ray mask, and initial wavelength solution generation. For LRIS wide slit, we entirely used a custom reduction pipeline due to the novel nature of the mode. Finally, for KCWI we largely relied on the KCWI Data Reduction Pipeline\footnote{\href{https://github.com/Keck-DataReductionPipelines/KCWI_DRP}{github.com/Keck-DataReductionPipelines/KCWI\_DRP}} (KCWI DRP) for initial individual frame reduction, with some custom final steps to combine 2D frames into a final 1D spectrum. Details for these three reduction procedures are given below.


\subsubsection{LRIS Narrow Slit} \label{Sec:Narrow}

In order to reduce data taken in the LRIS narrow slit mode over two globular clusters in each of three galaxies, we began by using {\tt PypeIt} to construct an initial sky model, cosmic ray mask, and wavelength solution for each frame. This served as the basis for our tailored reduction steps.

We improved the accuracy of these initial wavelength solutions in the following way. First, we measured the wavelengths of 19 sky lines, ranging from 7653.313 {\AA} to 8943.395 {\AA} (in air wavelengths), by removing the heliocentric correction applied by {\tt PypeIt} and fitting Gaussians to lines near the expected wavelengths. We then fit a fifth order polynomial to provide a more accurate wavelength solution. Finally, we applied these solutions using {\tt interp1d}, placing all frames on a common wavelength grid.

We then straightened the 2D spectra in the y direction and placed them on a common 2D grid. First, we sampled the spectrum at ten fixed wavelength intervals. Within these ten bins we fit flux across the position axis, with each pixel averaged across the wavelength bin, with a Gaussian centered at the peak of the trace of the brightest object (i.e., the brightest globular cluster). We then fit a polynomial to provide correcting shifts such that the peak of the fit trace at all bins was on the same pixel for all frames. Finally, we applied these solutions placing all frames on a common 2D grid with a straightened trace.

Next, we combined individual frames into a master frame, accounting for differences in transparency and seeing. We created a slit profile for each frame, summed across the wavelength direction, and simultaneously fit objects in these profiles with Gaussians to calculate the ratio of the integrated flux to the FWHM. The average ratio of each frame was used as the weight for that frame. The binary cosmic ray masks, reduced with the same correcting steps as above such that all pixels affected by cosmic rays following interpolation were masked, were here applied such that at each pixel in the master frame is the weighted average of all non-masked pixels.

Finally, we extracted a 1D spectrum for each object. We created a slit profile for the master frame, summing across the wavelength direction, and modeled objects in this slit profile with Gaussians in a simultaneous fit. For RCP32, 1D globular cluster spectra could then be extracted by performing a weighted sum, multiplying the flux at each position by the fit Gaussian profile and adding them up at each wavelength, so long as the fit profile met a threshold 15\% of the peak. For DF9 and LEDA 4014647, diffuse galaxy light overlapped with globular cluster spectra, contaminating one another. We therefore removed contaminating light to create a separate spectrum, both for the globular clusters and diffuse galaxy light, in the following way. A `cleaned' 1D spectrum for the galaxy was extracted using the same method as above (i.e., a sum using a fit Gaussian profile as weights, above a 15\% threshold) by simply masking positions in which globular clusters were present at a level $>$1\% of the diffuse galaxy light. This cleaned 1D spectrum, times the profile of the galaxy at each position, was then subtracted across the entire 2D spectrum creating a 2D spectrum with only globular clusters. We then re-fit these subtracted 2D spectra to obtain a new slit profile, and extracted `cleaned' 1D spectra for them as well, using the same sum weighted by fit profiles as in RCP32. To extract an error spectrum, we repeated the above steps starting with the square root of the sky counts, following propagation of error rules during weighted sums.


\subsubsection{LRIS Wide Slit} \label{Sec:Wide}

We developed a novel technique to measure radial velocities of diffuse, low luminosity galaxies with LRIS. We use a wide slit, of $5\arcsec$ or $7\arcsec$, to capture a much larger fraction of the galaxy's light than a standard $1\arcsec$ wide slit would. The price is a much-reduced spectral resolution, which typically means that absorption lines are more difficult to detect. We mitigate this by focusing on the Ca II H and K lines at 3968.469 and 3933.663 \AA (Air), as they are intrinsically broad. Furthermore, we use a high resolution grism (the 1200 lines mm$^{-1}$ grism), so that the spectral resolution is R$\approx$500 even after the factor of 5--7 degradation by the broad slit. This spectral resolution is well-matched to the intrinsic width of the Ca H and K lines, which means that there is little benefit from having a narrower slit. Two technical modifications have to be made: First, a custom slit mask had to be designed with the wide slit in the far right of the mask, so that the observed wavelength coverage extends far enough into the red. Second, the spectra are oversampled by a large factor and the wavelength coverage is small. The oversampling is dealt with by binning. The small wavelength coverage could have been an issue for wavelength calibration, as there are not many sky lines or arc lines near 4000\,\AA; however, as explained below, we find that we can use the Ca II H and K lines of the solar spectrum, as present in the zodiacal light. Our wide slit captures sufficient background light to enable this measurement for all our objects.

Due to the novel nature of LRIS wide slit mode observations, which aims to observe the broad Ca II H and K lines in a high resolution grating and binning heavily, reduction was performed with a custom pipeline. 

The bias subtraction in this case was non-trivial as the overscan regions showed clear evidence of a gradient. To model this (while ignoring cosmic rays and hot pixels), we took a median over the overscan region collapsing into 1D, taking a nine pixel rolling average to account for slow binary increases, and fitting with a ninth order polynomial (as was empirically, if somewhat arbitrarily, determined to be an effective model). This 1D fit was then expanded to the entire frame and subtracted. 

We then removed hot pixels and cosmic rays in the frames so that they were not smeared out during resampling and binning. This was achieved by first applying 2$\times$2 binning on the frames, such that cosmic rays with fuzzy edges stood out, then running {\tt L.A.Cosmic} \citep{2001PASP..113.1420V}.

Next, given that the data are oversampled by a factor of $\sim$10 in wavelength and we extract a spectrum over a wide trace, we binned the frame by 5 in wavelength and 2 in position. In order to keep track of counts, we summed rather than averaged while binning.

Due to how the grism projected light on the CCD, the spectra curved along the position axis and required rectification. Moreover, this curvature changed slightly across the slit, so we first modeled the curved shape of the slit edges across the detector, then fit a line to capture how the required corrections at each slit edge would change over the intervening pixels (this was an appropriate approximation given that the asymmetry in the curvature was small). To model the slit edges, we sampled them along the wavelength direction by averaging over slices with a width of fifty pixels and identifying the edges in the resulting slit profiles. The resulting shifts to straighten each edge were then extended across the entire detector with a linear fit. Given that the shifts here were calculated so that edges were straightened into the same pixel coordinates for each frame, this means that all the individual frames were then aligned as well.

We then created master flat fields and applied them to each frame. At this point in the reduction, individual flat field frames had already received the same corrections as outlined above. The flat fields showed strong gradients from blue to red due to the effective spectrum of the flat field lamp and from bottom to top due the location of the lamp relative to the light path. These were modeled and removed by dividing by low order polynomial fits in both directions. The flats were then averaged to generate a master flat and applied by dividing each science frame by this master flat.

Next we aligned the wavelengths of the frames, which required getting a reference background spectrum and measuring the tilts of lines down the slit. This 1D reference spectrum was generated by averaging over 25 pixels in the top portion of the spectrum found to be object free in all frames. Next, to identify the tilted background line locations down the slit, we sampled across the position, axis averaging over slices with a width of ten pixels. Finally, for each of these spectra we identified shifts required to make the background lines fall on the reference spectrum by calculating the chi-squared value for shifts over a grid from $-$9.9 to 9.9 pixels with a spacing of 0.1 pixels. As this allows for sub-pixel shifts (the line tilts can be gradual across frames, e.g. requiring a shift of 0 pixels at the bottom of the slit and 1$-$2 pixels at the top), we first re-gridded the spectra with 0.025 pixel increments. We then fit these shifts with a line and applied the corrections such that all frames were straightened and aligned on the same spatial and wavelength grid.

Wavelengths were calibrated by modeling this observed reference spectrum. The dominant features in the background spectrum are the solar Ca II H and K lines due to zodiacal light. A background model was created by combining a Cerro Paranal Sky Model (\citealt{2012A&A...543A..92N,2013A&A...560A..91J}; as accessed via the ESO Sky Model Calculator) with a solar spectrum to model the zodiacal light dominant here at high resolution. These spectra were given in vacuum wavelengths which we converted to air using the IAU standard. We then fit the observed reference spectrum with the combined model, using a third order polynomial to find the wavelength as a function of pixel location.

The background was subtracted using a model generated from portions of the slit without objects. Since the background varied over the position axis, this required fine tuning to identify object free regions throughout the slits, and most critically, near the galaxy traces themselves so that the background was correctly modeled there. The background was modeled with a line at each wavelength, and then subtracted.

Finally, to get a 1D spectrum from the 2D individual frames, as well as an error spectrum, we followed essentially the same process as in Section~\ref{Sec:Reduction} with the exception that all targets only required a single Gaussian. For DF5, the target taken with both 5$\arcsec$ and 7$\arcsec$ slits, two separate 1D spectra were generated and combination occurred at this stage. Since these utilized different reference spectra, combination on a common grid required an additional interpolation step.

As a sanity check to confirm our reduction of this novel observing mode, several other targets (not discussed in this paper) were reduced with this pipeline to compare to literature values, including  DF4. These tests generally were in excellent agreement within estimated uncertainties; in particular, the central value of our DF4 radial velocity measured in this way, 1430$^{+20}_{-10}$ km s$^{-1}$, is in agreement with the 1433 km s$^{-1}$ measurement by \citet{2023ApJ...957....6S}. We utilize the measurement by \citet{2023ApJ...957....6S} during the analytical portions of this work due to its lower associated uncertainty. 


\subsubsection{KCWI} \label{Sec:KCWI}

In order to reduce the IFU data cubes generated by our KCWI light bucket mode observations, we utilized the KCWI DRP pipeline.

For the two galaxies which did not include alternating sky offsets, Ta21-12000 and RCP26, we allowed the pipeline to proceed with automated sky subtraction which identifies areas of the sky that do not contain object flux. For DF7, which was observed with an alternating sky frame, sky subtraction was ignored. Instead, following an initial step converting vacuum to air wavelengths and masking bad wavelengths with known artifacts, the average of neighboring off-target sky frames was subtracted from on-target science frames. Note that for Ta21-12000 there was an additional strong artifact from switching instrument configurations, resulting in a significantly larger wavelength range being masked.

Obtaining a 1D spectrum involved a similar procedure as in the above LRIS modes, including a spatial Gaussian fit with the same 15\% inclusion threshold, with the exception that Gaussians were now in 2D. Moreover, for RCP26, a second Gaussian model for an overlapping galaxy had to be simultaneously employed.


\subsection{Radial Velocity Measurements} \label{Sec:Fit}

\begin{figure*}
    \setlength{\lineskip}{0pt}
    \centering
    \includegraphics[width=0.515539257981018\textwidth]{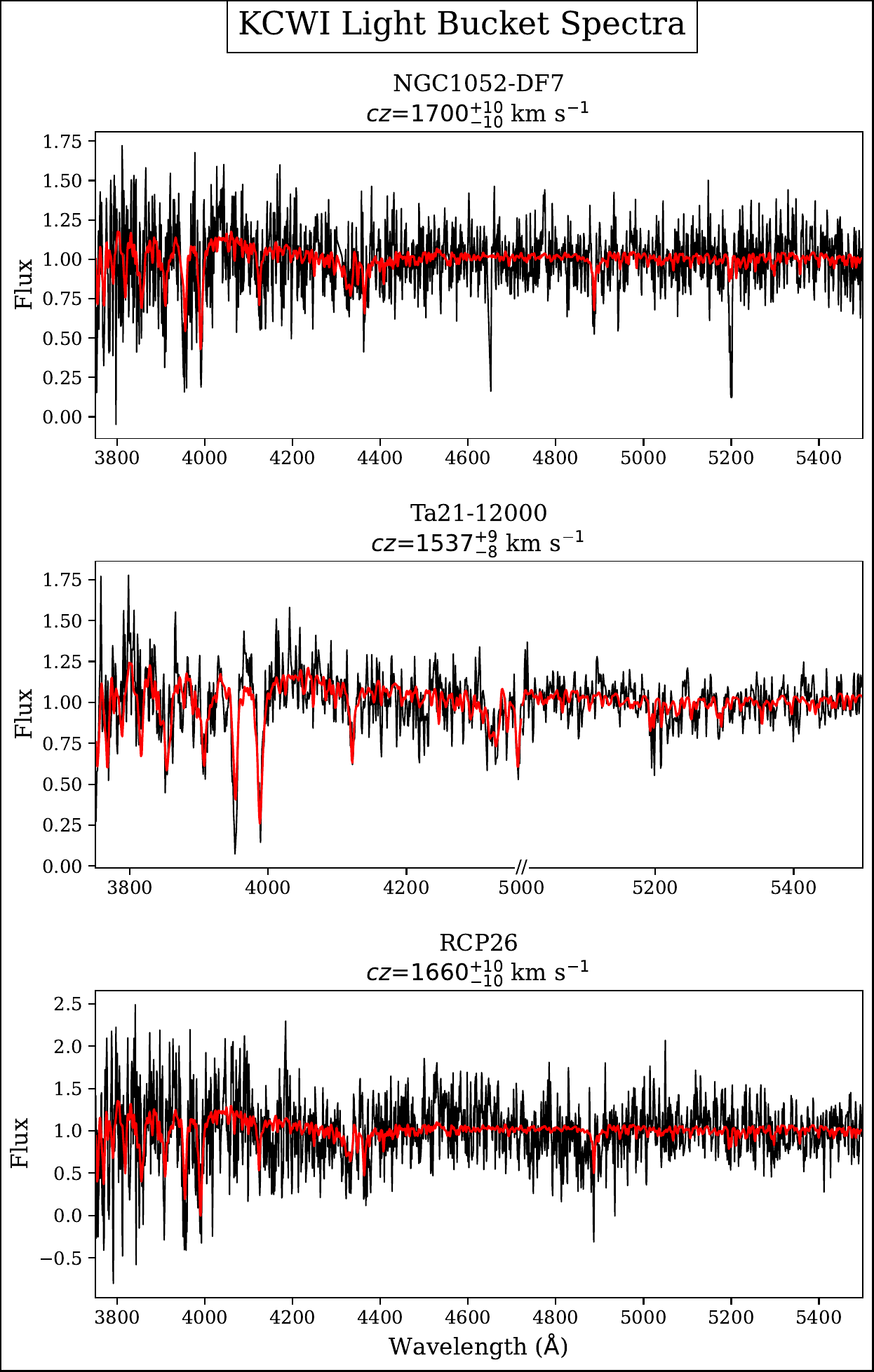}\includegraphics[width=0.474460742018982\textwidth]{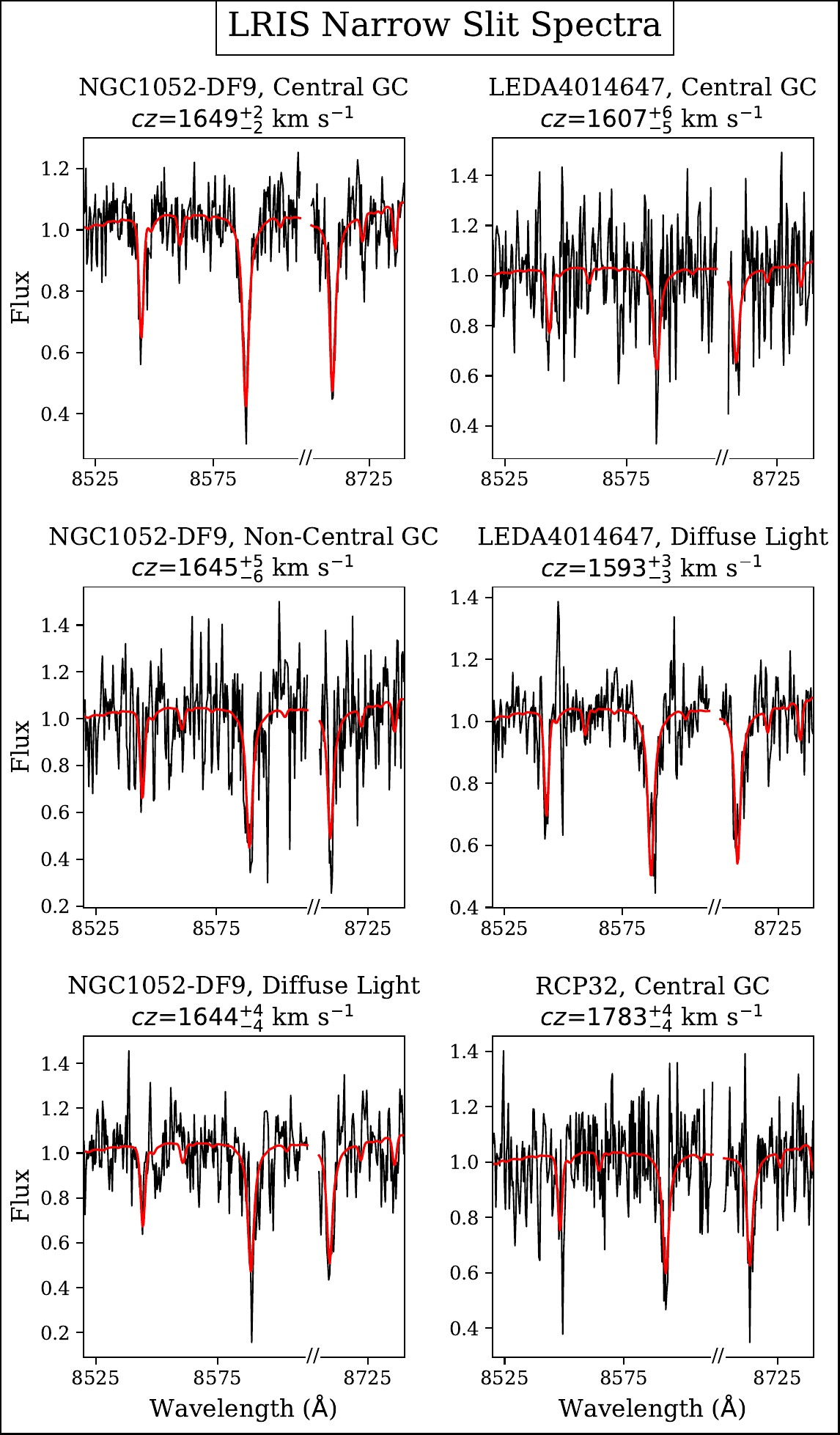}\\  \includegraphics[width=0.99\textwidth]{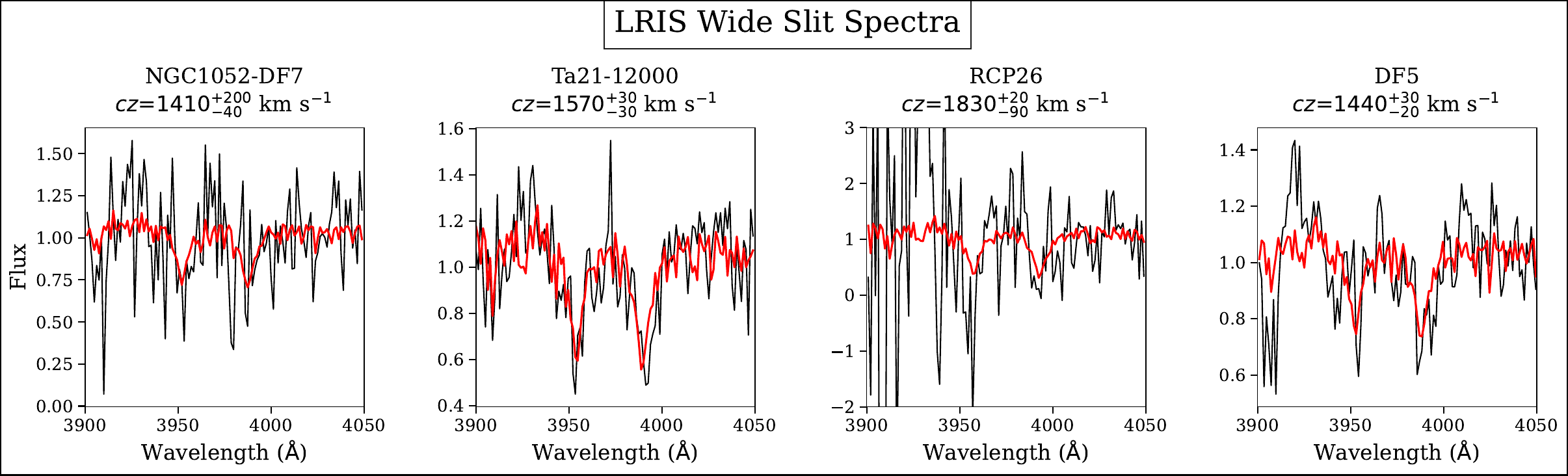}
    \caption{Extracted 1D spectra post-regularization (\textit{black}) and fit templates (\textit{red}), for each observing mode (\textit{top left: LRIS wide slit, top right: KCWI light bucket, bottom: LRIS narrow slit}). Spectra are given in air wavelengths. Galaxy and, for the narrow slit mode, object names as well as fit velocities for individual spectra are indicated in plot titles. \label{Fig:Spectra}}
\end{figure*}

In order to determine radial velocities, we fit 1D spectra using the {\tt emcee} Markov chain Monte Carlo (MCMC) sampling algorithm \citep{2013PASP..125..306F}. First, spectra are restricted to specific domains of interest: the calcium triplet region between 8520 {\AA} and 8740 {\AA} for LRIS narrow slit spectra, the calcium H and K Fraunhofer lines between 3900 {\AA} and 4050 {\AA} for LRIS wide slit spectra, and a range of lines (mostly Fe) between 3600 {\AA} and 5500 {\AA} for KCWI light bucket spectra (note that most of the region between the calcium triplet lines was masked due to strong sky lines, and is not shown in this work). They are then regularized by dividing by a polynomial based on the spectral range (2nd, 1st, and 5th, respectively). The spectra are modeled with simple stellar population templates with metallicities and ages near that of DF2 (${[\mathrm{M}/\mathrm{H}]\approx-1}$, ${\approx}8$ Gyr; \citealt{2019A&A...625A..77F,2022Natur.605..435V}). Finally, the velocity and an additive continuum offset are fit using one hundred walkers with one thousand samples generated. We verify that burn-in has occurred through visual inspection of the distribution of walker positions. We check the accuracy of our error spectrum by plotting the distribution of the fit residuals divided by estimated uncertainties, which we confirm has a standard deviation of $\sim$1. The ultimate radial velocity uncertainty is determined from 16th and 84th percentiles.

The best fitting model spectra for all three methods, along with inferred radial velocities for the individual spectra, are shown in Figure~\ref{Fig:Spectra}. We apply a heliocentric correction for LRIS narrow slit observations, which, though automatically applied by {\tt PypeIt}, had been removed for sky line re-analysis, and for LRIS wide slit observations, which our custom pipeline had yet to apply. This was not necessary for KCWI light bucket observations since the KCWI DRP automatically applies the correction. Besides RCP32 and DF5, the remaining five targets all have two to three radial velocity measurements either for multiple objects or with multiple observing modes. To interpret the observed galaxy velocities with one combined value, we take a uncertainty-weighted average of either individual object velocities or the velocities from the various modes.\footnote{We note that the second RCP32 candidate globular cluster is, in fact, a star at a radial velocity of 0 km s$^{-1}$, and the second LEDA 4014647 candidate, after subtracting off the diffuse light of the galaxy, had 15$\times$ less flux in the slit than the other globular cluster and spectrum was not of sufficient quality to determine a reliable radial velocity.} These velocities, as well as, for reference, the velocities of DF2 and DF4 and the relative positions of all galaxies along the trail, are given in Table~\ref{Table:RVs}.

\begin{deluxetable}{ccc}
\tablecaption{Final radial velocity measurements for our seven targets, DF2 \citep{2019ApJ...874L..12D}, and DF4 \citep{2023ApJ...957....6S}, along with the positions of galaxies too faint for redshift determination.\label{Table:RVs}}
\tablewidth{0pt}
\tablehead{
\colhead{Source} & \colhead{[ $\Delta x$, $\Delta y$ ]\tablenotemark{a}} & \colhead{Radial Velocity} \\ & \colhead{(arcmin)} & \colhead{(km s$^{-1}$)}
}
\startdata
RCP32       & [   29.8,    5.8 ] & $1783^{+4}_{-4}$       \\
NGC 1052-DF2 & [   20.8,    0.0 ] & $1805^{+1}_{-1}$       \\
RCP28       & [   18.2,    2.4 ] & $-$                    \\
RCP26       & [    9.7,    1.9 ] & $1670^{+10}_{-10}$     \\
RCP21       & [ $-$0.9,    0.7 ] & $-$                    \\
NGC 1052-DF9 & [ $-$6.2, $-$0.8 ] & $1648^{+1}_{-1}$       \\
RCP17       & [ $-$9.9, $-$1.4 ] & $-$                    \\
Ta21-12000  & [ $-$12.5, $-$3.6 ] & $1540^{+9}_{-8}$       \\
NGC 1052-DF5 & [ $-$20.8, $-$1.6 ] & $1440^{+30}_{-20}$     \\
NGC 1052-DF4 & [ $-$20.8,    0.0 ] & $1433.3^{+0.3}_{-0.4}$ \\
LEDA 4014647 & [ $-$28.5,    0.9 ] & $1596^{+3}_{-2}$       \\
NGC 1052-DF7 & [ $-$35.9,    5.7 ] & $1690^{+10}_{-10}$     \\
\enddata
\tablenotetext{a}{Positions are given in coordinates relative to the line connecting DF2 and DF4 (above or below this line on the y-axis, and along this line relative to the midpoint between DF2 and DF4 on the x-axis), i.e. approximately the projected collision axis.}
\end{deluxetable}


\section{Results} \label{Sec:Results}

Following radial velocity determination, we may now compare to expectations from the bullet dwarf collision scenario. We both compare to the kinematic trend expected from DF2 and DF4 alone, and consider them together with our new measurements.

\begin{figure*}
    \centering
    \includegraphics[width=0.47\textwidth]{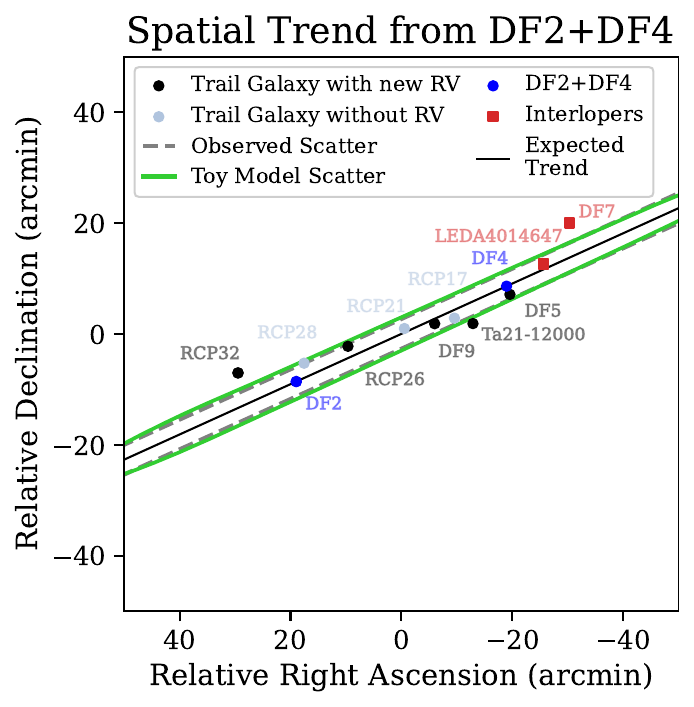}\quad\includegraphics[width=0.48831168831\textwidth]{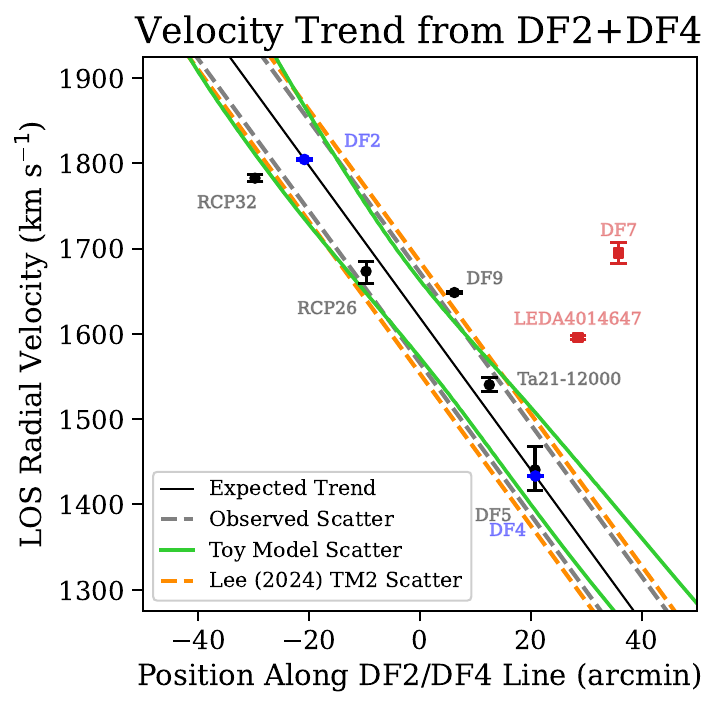} 
    
    \vspace*{-3mm}
    \caption{The spatial positions of galaxies along the trail (\textit{left panel}) and the radial velocities measured in this work (\textit{right panel}), as compared to the trend (\textit{black lines}) expected from collisional formation based on DF2 and DF4 alone. Positions and and radial velocities, with associated uncertainties, are indicated for DF2 and DF4 (\textit{blue circles}), trail galaxies which follow DF2 and DF4's expected kinematic trend (\textit{black circles}), and interlopers which do not appear to do so (\textit{red squares}). Positions for yet fainter galaxies still without radial velocities are also indicated (\textit{grey circles}). The observed rms relative to the both trends (\textit{grey dashed lines}) is calculated after subtracting the expected positions (\textit{left}) and velocities (\textit{right}) as based on DF2 and DF4, i.e. the scatter relative to the expected trends. These are compared to simulation, both \citet[with the rms scatter for their TM2 run calculated in the same manner as for the observed trend, using the DF2 and DF4 equivalent galaxies for reference]{2024ApJ...966...72L} and to the toy model in Appendix~\ref{Sec:ToyModel} (with the density contours containing 68.27\% of 100,000 test particles in a simulated bullet dwarf collision). In the right panel, positions are given as along the line connecting DF2 and DF4, i.e. the expected spatial trend given by the black line on the left panel.\label{Fig:Trend}}
\end{figure*}
\begin{figure*}
    \centering
    \vspace*{-5mm}
    \includegraphics[width=0.47\textwidth]{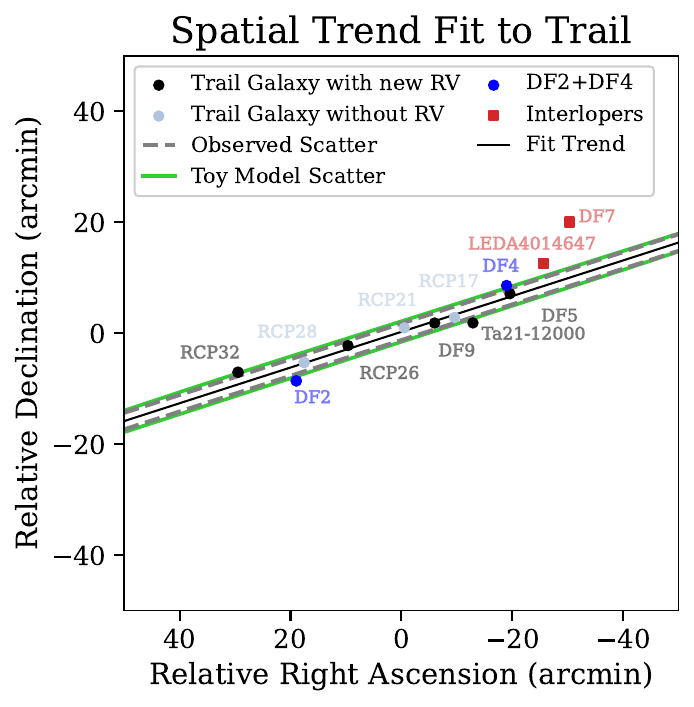}\quad\includegraphics[width=0.480871886120996\textwidth]{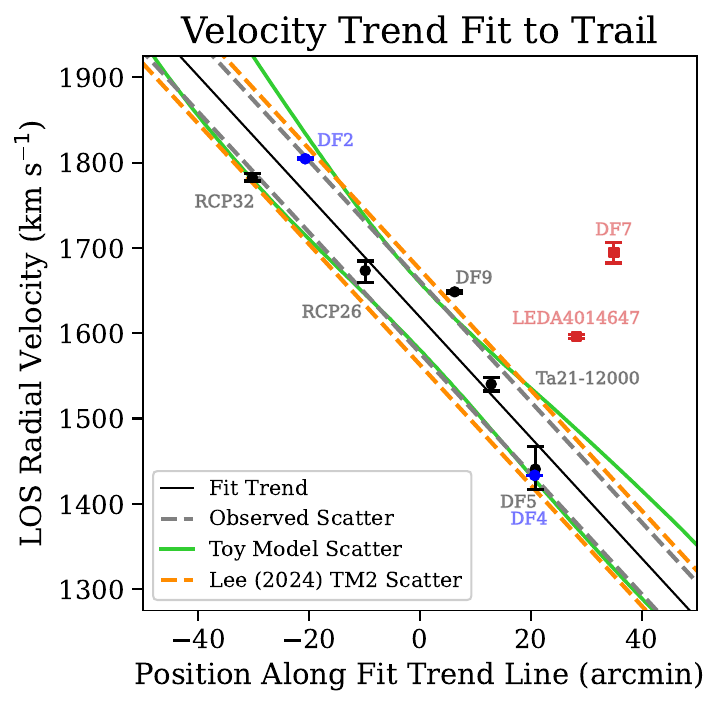} 
    
    \vspace*{-3mm}
    \caption{Same as Figure~\ref{Fig:Trend} with trend lines fit to all trail galaxy positions and velocities (excluding interlopers). The observed rms scatter and the rms scatter from \citet[TM2]{2024ApJ...966...72L} are recalculated to be relative to these fit trend lines, and the toy model is adjusted to match the tighter trend (see Appendix~\ref{Sec:ToyModel}).\label{Fig:TrendAdjusted}}
\end{figure*}

\subsection{Radial Velocities} \label{Sec:RVs}

Radial velocity measurement results for our various targets, given in Table~\ref{Table:RVs}, are plotted in Figure~\ref{Fig:Trend} on top of the velocity trend expectation based on DF2 and DF4. Five of the seven galaxies follow the expected trend: RCP32, RCP26, DF9, Ta21-12000, and DF5. Two on the western end are $>$100 km s$^{-1}$ above their expect velocities from the trend, LEDA 4014647 and DF7.

\citet{2022Natur.605..435V} determined that 2$\pm$2 of the 12 galaxies along the trail were `normal' members of the NGC 1052 that fell along the trail in projection on the two-dimensional sky by chance. In other words, it is unlikely that all 12 galaxies are genuine trail members. \citet{2022Natur.605..435V} identified LEDA 4014647 as a candidate interloper based on its high relative luminosity and compact size. Although \citet{2022Natur.605..435V} did not anticipate DF7 to be an interloper, it is spatially offset from the rest of the trail once LEDA 4014647 is removed, being $960\arcsec$ from its nearest neighbor DF4, as compared to RCP32 being $631\arcsec$ from DF2 and an average nearest neighbor at $272\arcsec$. 

In order to determine such interlopers on a quantitative basis, in  Appendix~\ref{Sec:ToyModel} we run a toy model of the collision. We utilize 100,000 test particles, and explore both their spatial distribution as projected on the sky and their radial velocity distribution along the line of sight. Because the trail is so tightly aligned on the sky, we find that it would be highly unlikely for LEDA 4014647 and DF7 to be genuine members of the trail.

In Figure~\ref{Fig:TrendAdjusted}, we instead plot using a trend fitting all galaxies, including both prior DF2 and DF4 measurements as well as the new trail galaxy velocity results (excluding the interlopers LEDA 4014647 and DF7), rather than just the trend purely expected from DF2 and DF4 alone. We adjust our coordinate system to be relative to the entire trail, with a line fit to all galaxies except LEDA 4014647 and DF7. Both Figures~\ref{Fig:Trend} and~\ref{Fig:TrendAdjusted} include the $1\sigma$ contours for the distribution of test particles in the toy model explored in Appendix~\ref{Sec:ToyModel}, where the distribution of transverse velocities and accelerations has been tuned to match the observed rms scatter in each case.

We note that both the expected velocity trend based on DF2 and DF4 alone and the fit trend in Figure~\ref{Fig:TrendAdjusted} are highly distinct from the velocity distribution of the group, with a mean velocity of 1435 km s$^{-1}$ and a line-of-sight velocity dispersion of 115 km s$^{-1}$ (\citealt{2022Natur.605..435V}; we do not utilize the line of sight dispersion of 143 km s$^{-1}$ from \citealt{2017ApJ...843...16K} as it contains only half of the galaxies with current spectroscopically confirmed redshifts). The eastern side of the trail lying at the expected radial velocities would place them outside of the group.

We may determine the `scatter' relative to both the DF2$-$DF4 trend as well as the trend fit to the trail galaxies by calculating the rms of the deviations between the galaxies' observed radial velocities and their radial velocities expected for the trend. For the DF2$-$DF4 trend this is 53 km s$^{-1}$, and for the trend fit to the trail galaxies this is 42 km s$^{-1}$.

\subsection{Comparison to Simulation} \label{Sec:Simulation}

\citet{2024ApJ...966...72L} conducts a series of impressive gravitohydrodynamic simulations to create close analogs to the bullet dwarf trail. In Figures~\ref{Fig:Trend} and~\ref{Fig:TrendAdjusted} we compare the observed `scatter' relative to the trend to the results of \citet{2024ApJ...966...72L}. Table~4 of \citet{2024ApJ...966...72L} lists the positions and velocities of galaxies in three simulations. Importantly, these each include DF2 and DF4 equivalents. We may therefore calculate both a DF2-equivalent$-$DF4-equivalent velocity trend and a velocity trend, just as we did for the observed trail, and compare the rms scatter relative to those trends to our own findings. Notably, however, the rms itself must be considered relative to the velocity difference between DF2 and DF4, as a more steep trend will naturally have a larger rms.

Dividing the observed rms scatter found in our work by 371 km s$^{-1}$, the velocity difference between DF2 and DF4, we find a relative rms scatter of 14\% from the DF2$-$DF4 trend and 11\% from the trend fit to the trail galaxies. Doing the same for the TM1, TM2, and TM3 simulations of \citet{2024ApJ...966...72L}, we find relative scatters of 59\%, 18\%, and 56\% from the DF2-equivalent$-$DF4-equivalent trends and 28\%, 15\%, and 24\% from the trend fit to the trail galaxies. The one closest to our results, TM2, is included in Figures~\ref{Fig:Trend} and~\ref{Fig:TrendAdjusted}.

It is naturally compelling that one of these three simulations has so similar scatter to that we observe. Still, the majority of simulations in Table~4 of \citet{2024ApJ...966...72L} have much higher scatter than what we observe. It is possible that TM2 simply describes the observed trail better. However, both TM1 and TM3's rms are driven up primarily by just one or two galaxies which are $>$200 km s$^{-1}$ off from their expected velocity. This may lead one to believe that DF7 or LEDA 4014647 may be a true dark matter deficient galaxy which is a member of the trail, and not, in fact, an interloper. Indeed, if we include both DF7 and LEDA 4014647, $rms/\langle v_{\mathrm{DF2}} - v_{\mathrm{DF4}} \rangle$ would increase from 11\% to 43\%, closer to that seen in TM1 and TM3 of \citet{2024ApJ...966...72L}. Even so, it is important to consider that the true spatial distribution of the galaxies above and below the axis of the trail is very tight, which means that the scatter of velocities cannot be very high. It is for this reason that we constructed a toy model which, though far simpler than the simulation \citet{2024ApJ...966...72L}, includes a larger number of galaxies (test particles) to explore the resulting distribution and, importantly, compares to the observed two-dimensional distribution as projected onto the sky. We conclude that the tight alignment of the trail on the sky suggests that DF7 and LEDA 4014647 are interlopers, rather than trail members with randomly high velocities.

\subsection{Probability Analysis} \label{Sec:Prob}

\begin{figure}
    \centering
    \includegraphics[width=0.45\textwidth]{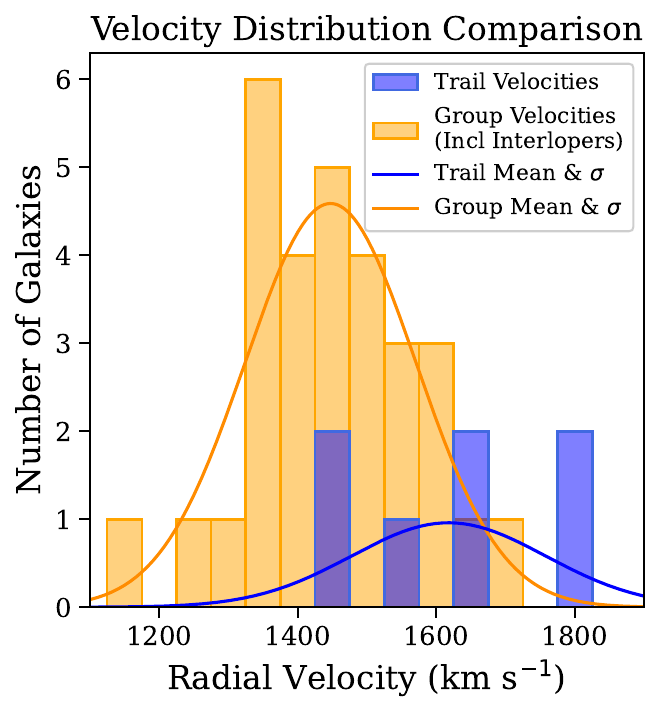}
    \caption{A comparison between the velocity distributions of the NGC 1052 group and the `bullet dwarf' trail. Note that the interlopers LEDA 4014647 and DF7 have been included in the group distribution and not the trail. A KS test indicates the trail and the group are from different velocity distributions at the 7\% level.\label{Fig:Distribution_Comparison}}
\end{figure}

Here we analyze the probabilistic significance of our findings. A few individual results are themselves notable; in particular, RCP32, like DF2, is a 3$\sigma$ outlier compared to the NGC 1052 group's velocity distribution. In addition, the chance of having DF4 and DF5, two diffuse galaxies, randomly appearing within 1.6 arcmin of each other and having radial velocities within 7.4 km s$^{-1}$ is very low if one considers them to be random objects of the group.\footnote{Taking one million random draws of the group's diffuse galaxies, modeling the positions of 42 galaxies of \citealt{2021A&A...656A..44R} with a $\sigma_r = 1.23^\circ$ circular Gaussian (this includes the trail galaxies as well, inflating the probability) and assigning them random velocities based on the group's mean and dispersion, we find that two diffuse galaxies would randomly lie so close to each other with such similar velocities only 1.7\% of the time (a figure which accounts for the +27 km s$^{-1}$ uncertainty in our DF5 velocity measurement).} However, it is not necessarily uncommon to find such close gravitationally bound dwarf galaxies -- see Section~\ref{Sec:DF5} for further discussion.

Considering the trail as a whole, there are several remarkable aspects are, independently, unlikely: First, that this tight geometric alignment of low surface brightness galaxies with DF2 and DF4 appeared to begin with. Second, that the trail galaxies have much higher velocities compared to the group, including several high-$\sigma$ outliers. And finally, that 5 of the 7 targets followed the specific predicted trend from DF2 and DF4. Analyzing all of these factors together does not merit a very informative result - in a computationally feasible number of random draws of galaxies from the group's velocity and density profile, one would not expect to find \textit{any} equivalent trails which include DF2, DF4, and galaxies with high velocities like RCP32 following the inferred trend based on the positions of DF2 and DF4. Thus, we instead consider each factor independently, to better provide insight into the unique aspects of our findings.

First, we consider the tight linear alignment of the trail galaxies. The low geometric odds of such a compact linear alignment has already been explored by \citet{2022Natur.605..435V}. To summarize, such a spatial arrangement has a high significance because it stands out relative to the density of the group (i.e., the same trail in a denser cluster would have far lower significance). Specifically, the odds of such an arrangement occurring by chance in the NGC 1052 group and containing both DF2 and DF4 are just 0.6\% (\citealt{2022Natur.605..435V}). This analysis did not include DF9 as it was not part of the sample population due to its bright central cluster (nor LEDA 4014647 due to its brightness and compactness), though it did include DF7 which was the most distant from the collision axis and an interloper. By replacing DF7 with DF9 thereby making the trail even tighter, a similar test would show an even higher significance.

Second, we compare the velocities of the trail galaxies to the NGC 1052 group. That is, putting aside the low odds that the galaxies would randomly have this geometric arrangement to begin with, known before this work, here we can consider separately the probability that the velocities of the trail galaxies arose randomly from the group. The fact that the trail contains so many high-$\sigma$ outliers compared to the group, including RCP32 and DF2 at 3$\sigma$ above the group mean, and that all but one of the trail galaxies have velocities higher than the mean of the group, is already compelling evidence that the galaxies did not randomly arise from the group. However, given that LEDA 4014647 and DF7 are now identified as interlopers on the trail, we should consider their inclusion as part of the NGC 1052 group. Together with the 28 galaxies in group with previously known velocities (\citealt{2021A&A...656A..44R}; with Table 2 modified as in \citealt{2022Natur.605..435V}), the adjusted mean velocity and line-of-sight velocity dispersion are 1447 km s$^{-1}$ and 123 km s$^{-1}$, respectively. In this case, RCP32 and DF2 are 2.9$\sigma$ and 2.7$\sigma$ outliers. A summary of our statistical tests based on whether LEDA 4014647 and DF7 are included in the group or trail is given in Table~\ref{tab:stat_tests}. In Figure~\ref{Fig:Distribution_Comparison}, we give the distribution for both the trail, excluding interlopers (i.e. RCP32, DF2, RCP26, DF9, Ta21-12000, DF5, and DF4) and the group (i.e., LEDA 4014647, DF7, and the 28 galaxies in group with known velocities). Performing a Kolmogorov–Smirnov (KS) test for the two samples, we find a KS test statistic of $D=0.515$ and a P-value of $0.07$. We infer that the hypothesis that the two samples are drawn from the same underlying distribution is unlikely, though not quite formally ruled out at the 0.05 level.\footnote{Note that, while we include the interlopers LEDA 4014647 and DF7 in the group distribution for the KS test, it would be an incorrect application of the two sample KS test to include the remaining trail galaxies in the group distribution for the purposes of the KS test. By including the trail galaxies with the group population we would not be comparing the independent distributions, which would bias the result and change the meaning of the test statistic.} It is, therefore, unlikely that the trail galaxies arose randomly from the group.

\begin{deluxetable}{lcc}
\tablecaption{Comparison of statistical results assuming LEDA 4014647 and DF7 are included either on the trail or in the NGC 1052 Group\label{tab:stat_tests}}
\tablehead{
\colhead{Statistical Test} & 
\colhead{On the Trail} & 
\colhead{In the Group}
}
\startdata
Number of $>$2$\sigma$ Outliers & 4/7 & 3/5 \\
Number of $>$3$\sigma$ Outliers & 2/7 & 0/5 \\
KS Test P-Value & 0.005 & 0.07 \\
\enddata
\end{deluxetable}

Finally, we assess the odds that the galaxies' positions and velocities line up with the trend expected from DF2 and DF4. To quantify the significance of this finding, we first calculate the average deviation between the measured velocities and the expected velocities (based on the DF2--DF4 trend) for the five non-interloper galaxies, thereby establishing a baseline for comparison. Then, to test how often a similarly tight arrangement would occur by chance, we randomly draw velocities for each of the seven galaxies from a uniform distribution.\footnote{If we restrict the random draws to the observed velocity range, the random velocity drawn for DF7 can never be on the trend since it's expected velocity is below 1433.3 km s$^{-1}$. This would slightly inflate the apparent significance of our findings, giving a result that the arrangement would have randomly occurred 0.5\% of the time (i.e. even more significant than our actual result, by a factor of 4). To avoid this bias, we instead draw from a broader velocity range that includes all expected velocities from DF7 to RCP32. We also considered a velocity bounds 10\% above and below the range of expected velocities to account for boundaries, but found that this did not significantly alter our final result.} Then we select the five closest to the expected trend, tossing out the two furthest outliers as interlopers to mirror our treatment of LEDA 4014647 and DF7. By comparing the average velocity deviation of the five closest galaxies to the trend (that is, the five galaxies with randomly drawn velocities which are closest to the velocities expected from the DF2--DF4 trend for said galaxies) to the average deviation we actually observed for our five non-interloper galaxies, we can see if the random draw aligns with the expected trend as well as our true results. Ultimately, after repeating this procedure one million times, we find that  98\% of iterations had average velocity deviations greater than or equal to the observed average deviation. That is, in one million iterations such a tight arrangement of the 5 closest galaxies to the DF2--DF4 trend randomly occurs only 2\% of the time. We may conclude that this tight alignment with the expected trend is unlikely to be a chance occurrence.

Taken together, it is highly unlikely (a 0.6\% chance) that the entire trail is a chance alignment, it is unlikely (a KS test P-value of 0.07) that the observed radial velocities arose randomly from the group, and it is highly unlikely (a 2\% chance) that the radial velocities line up so well with the predicted trend from DF2 and DF4 by chance.


\section{Discussion} \label{Sec:Discussion}

The radial velocities observed for our targets in Section~\ref{Sec:Results} match expectations for the trail galaxies well. Several important implications of this discovery merit consideration.


\subsection{Velocity Trend Implications} \label{Sec:Trend}

Our observations imply that the trail, already with a high geometric significance, is a real, unique, kinematically connected structure. The trail is distinct from the NGC 1052 group, separated by several hundred km s$^{-1}$ at the eastern end with multiple high-$\sigma$ outliers in our relatively small sample. Thus, these galaxies are participating in the same unique motion, beyond the viral radius (which is 390 kpc for NGC 1052, whereas the trail stretches over $\approx$2 Mpc; \citealt{2019MNRAS.489.3665F,2022Natur.605..435V}), with the possible exception of the western end of the trail, at velocities so high that they are not as strongly influenced by the NGC 1052 group. These findings echo those of \citet{2024ApJ...966...72L} and explain why such a coherent trend can exist rather than having normal orbital velocities virialized within the group. It is likely that such a collision would have initially occurred outside of the virial radius, as required in simulations by \citet{2024ApJ...966...72L}, thereby explaining why the globular clusters of DF2 and DF4 were not stripped by NGC 1052 (as would have been potential concern if the galaxies were closer in; \citealt{2022ApJ...940L..46O}).

\begin{figure*}
    \centering
    \includegraphics[width=0.99\textwidth]{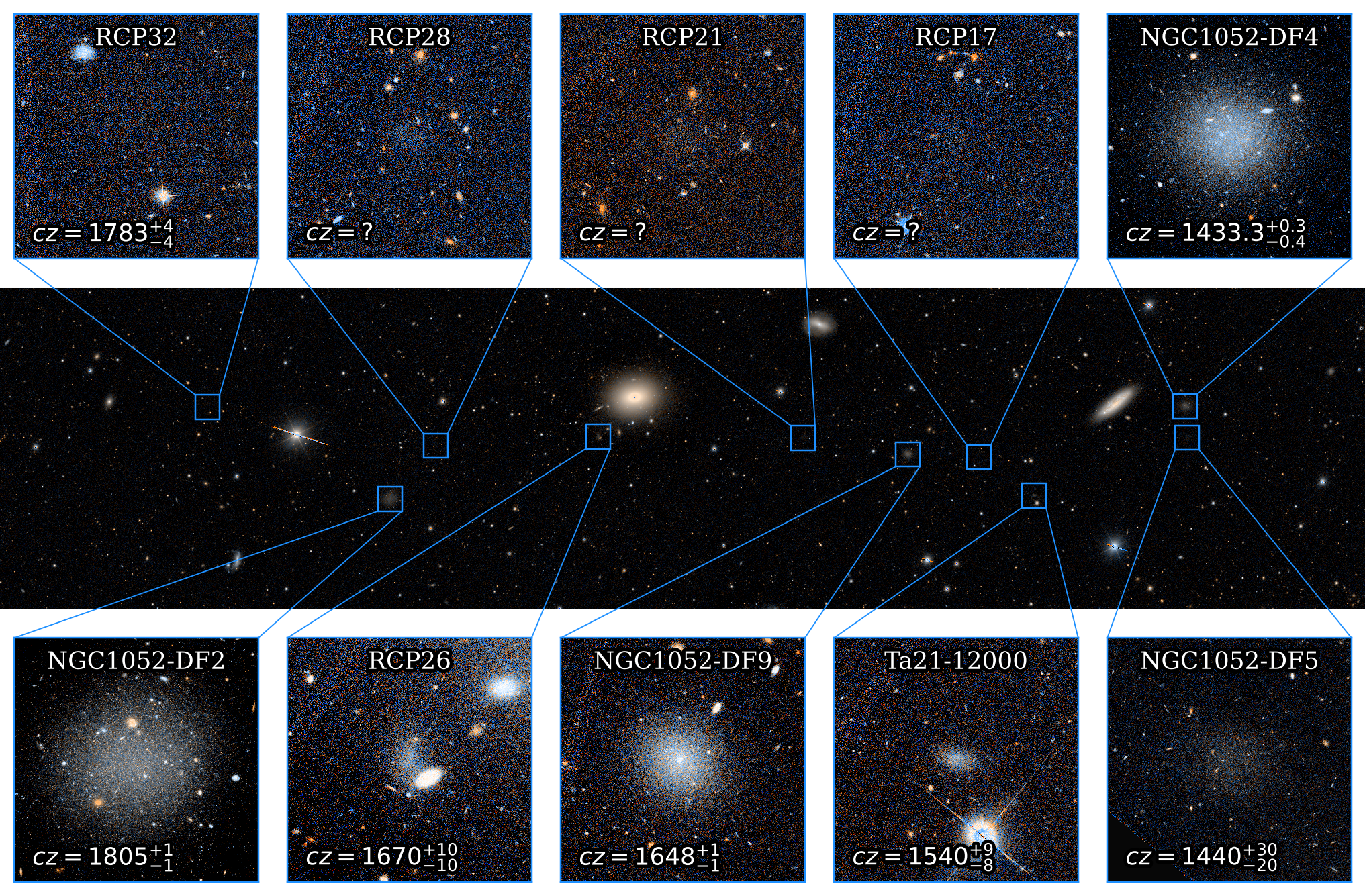}
    \caption{The trail as it now stands, with \textit{HST} cutouts of each member along with measured radial velocities over a DECaLS image of the field.\label{Fig:Final}}
\end{figure*}

Two of the seven measured radial velocities, those of LEDA 4014647 and DF7, are significantly more different from their expected velocities as compared to the rest of the trail. One possibility is that these are interlopers, chance projections which are not kinematically connected to the trail. \citet{2022Natur.605..435V} expected 2$\pm$2 galaxies on the line would be interlopers, and suggested LEDA 4014647 may be one based on its compact size and high surface brightness (moreover, DF7, upon removing LEDA 4014647 is highly spatially separated from the trail). However, LEDA 4014647 and DF7's velocities are also high compared to the group. A second possibility is that the galaxies are true members of the trail which happen to be offset from it. Indeed, \citet{2024ApJ...966...72L} found that it was not unusual for one or two galaxies on the trail to be $\approx$200 km s$^{-1}$ offset from their expected velocities, and, in fact, including all seven measured velocities in the observed scatter (i.e. assuming they are all trail members) still leaves our trend tighter than many of their simulated runs, rather than being tighter than all but one. A final possibility is that the two galaxies area actually the progenitors which retained their dark matter. \citet{2024ApJ...966...72L} implies that progenitors can have significantly different velocities as well, which would also be consistent with our findings for LEDA 4014647 and DF7.

Ultimately, this may mean that the dark matter free galaxy trail itself is shorter and tighter than initially thought, with DF4 and DF5 near the end. The final trail as it currently stands, including the trail galaxies' velocity measurements, is given in Figure~\ref{Fig:Final} (with \textit{HST} imaging of DF2 taken from program 15851, PI: van Dokkum).

\textbf{The observed velocity trend was a key prediction of the bullet dwarf collision scenario. It is a specific trend based on the velocities of DF2 and DF4, which the galaxies on the trail follow to a remarkable degree, which would have occurred by chance less than 2\% of the time. We set out to test the hypothesis that the galaxies would follow such a trend, and this hypothesis is now confirmed.}


\subsection{DF5 Velocity Consistent with DF4} \label{Sec:DF5}

\begin{figure}
    \centering
    \includegraphics[width=0.45\textwidth]{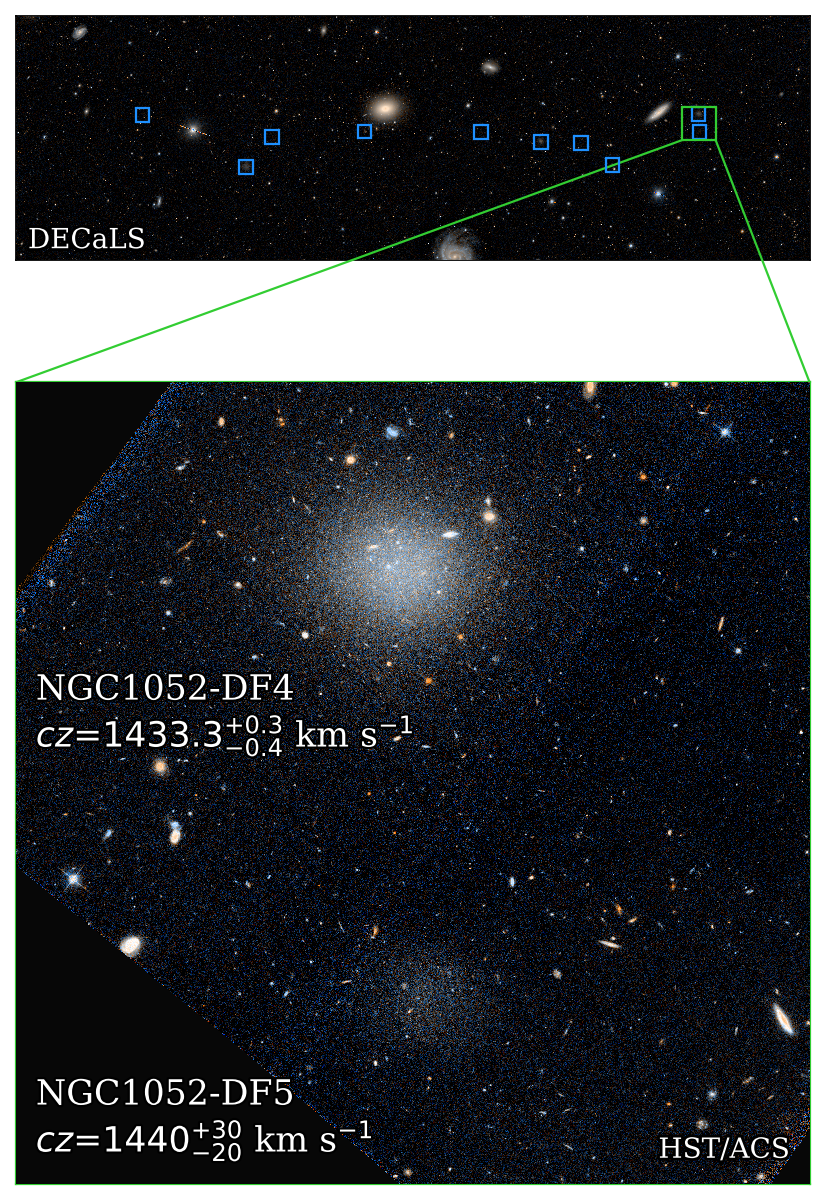}
    \caption{A demonstration of the similarities between DF4 and DF5. Both diffuse galaxies, which appear right next to each other on the sky, in the same deep HST pointing, and have nearly the same radial velocity.\label{Fig:DF45}}
\end{figure}

The bullet model predicts that two galaxies lying at the same position on the trail should have the same velocity. As we see in Figure~\ref{Fig:DF45}, the two similarly large, diffuse galaxies DF4 and DF5 lie right next to each other on the sky. In fact, when projected onto a line fitting the trail galaxies, they lie at almost the exact same position. Thus, it is highly interesting in this context that the measured radial velocity of DF5 is nearly the same as that of prior DF4 measurements \citep{2019ApJ...874L...5V,2021ApJ...909..179S}, with the newest DF4 measurement of \citet{2023ApJ...957....6S} lying well within our error bars. Indeed, multiple pairs of galaxies produced within the high speed collision simulation of \citet[see Figures 7 and 8]{2024ApJ...966...72L} have relative positions and velocities similarly close together, highly reminiscent of DF4 and DF5. We propose that these physical and dynamical similarities are due to formation via a common origin, in this case the bullet collision.

However, it is important to note that finding a bound pair of dwarf galaxies is not necessarily uncommon since it is a natural consequence of gravitational attraction. Examples include NGC 147 and NGC 185 in the Local Group, which appear to lie within 9 km s$^{-s}$ and 13 kpc of each other (\citealt{2013MNRAS.430..971W}; as projected on the sky at the distance of M31; \citealt{2012ApJ...745..156R}) similar to DF4 and DF5 separated by 7 km s$^{-1}$ and 9 kpc \citep{2020ApJ...895L...4D}. While the true 3D distance between NGC 147 and NGC 185 may be considerably larger \citep{2013MNRAS.430..971W}, this may be true for DF4 and DF5 as well (a difference too small for even the most accurate distance methods to discern). In any case, it is possible, even under the bullet scenario, that DF5 may have been gravitationally captured by DF4.


\subsection{Formation Mechanism} \label{Sec:Formation}

While this kinematic trend is a specific, original prediction of the bullet dwarf model, we may also consider the implied radial velocity trends for alternative formation scenarios for dark matter free galaxies. Such alternative models include gas flung out by quasars \citep{1998MNRAS.298..577N}, tidal dwarfs formed along tidal features created by a merger of massive galaxies (such as those from \citealt{2019A&A...628A..60F}), and tidal stripping of galaxies formed in normal galaxy formation channels \citep{2018MNRAS.480L.106O,2020PhRvL.125k1105Y,2021MNRAS.501..693M,2021MNRAS.502.1785J,2021MNRAS.503.1233O,2022MNRAS.510.2724O,2022NatAs...6..496M,2024arXiv240311403Z}. Though the NGC 1052 trail before its discovery was only predicted by collisional models \citep{2020ApJ...899...25S}, one can imagine exotic versions of all three alternative models that could form a trail of dark matter deficient galaxies (for tidal stripping, this might be explained by an infalling stream of galaxies all being stripped, similar to the linear substructures seen by \citealt{2023Natur.623..296W}).

Besides a bullet dwarf collision, alternative formation models for DF2 and DF4, and the associated trail of diffuse galaxies, already had difficulty explaining all the unusual properties of the galaxies prior to our work. In particular, DF2 and DF4's over-luminous, monochromatic globular clusters, the unique aspect of the galaxies which first made the galaxies stand out, point to a highly unique formation channel. Tidal stripping would not be able to explain the galaxies' bright globular clusters. Moreover, tidal stripping of more massive, metal rich normal galaxies which could not explain DF2 and DF4’s low metallicities; that is, given that the galaxies would have lost a large mass fraction and should a high metallicity for their mass \citep{2022NatAs...6..496M}. As for formation from gas expelled or excited by active galactic nuclei (AGN), few examples, or even simulations, exist for comparison (see, e.g., discussion in \citealt{2020MNRAS.499.4940Z}), so it is unclear how the galaxies produced would compare to the trail. On the other hand, many examples of tidal dwarfs forming along tidal features exist in nature, though they do not host the unusual globular clusters seen in DF2 and DF4. Moreover, the observed range of radial velocities along the trail, spanning nearly 400 km s$^{-1}$, is considerably larger than that which would be expected along the length of a tidal tail. Indeed, collisional formation remains the only robust explanatory model of this aspect shown in simulation \citep{2019MNRAS.488L..24S,2021ApJ...917L..15L}. Ultimately, only a high-speed collision can thus far explain all the unusual properties of the trail galaxies.

Our findings match the exact prediction from collisional formation, a specific kinematic trend based on the radial velocities of DF2 and DF4. Moreover, our findings set the trail apart from the group and mark it as a highly unique, connected system. Indeed, the similarities between DF2 and DF4, and now the spatially and kinematically connected trail, point to an origin in a common event. Still, it is important to compare the expected velocity trend to those from other formation scenarios.

If the galaxies formed from material flung out from an active galactic nucleus, or formed as tidal dwarfs from tidal features flung out in galaxy mergers, one might expect that they should still be kinematically connected to their host system, NGC 1052. That is because NGC 1052, a massive galaxy hosting a bright AGN, would be the most likely source of such AGN outflows, or the most likely merged product of tidal dwarf formation. Under this assumption, these two models are not consistent with the observed trend since they would predict a kinematic connection with NGC 1052, whereas NGC 1052 is instead offset from the trend based on its position relative to the trail by $\approx$200 km s$^{-1}$ (or $\sim$2$\sigma_{group}$, i.e. a larger difference than the dispersion of the group). Whereas the bullet dwarf model naturally explains this, with a collision occurring some distance away from NGC 1052, these alternative models would require some mechanism to decouple the host system from the dark matter free products in phase space. However, it is possible that there was an original progenitor or merging pair of progenitors besides NGC 1052 which may have kinematically decoupled from the trail galaxies due to the progenitor's massive dark matter halo, experiencing far greater dynamical friction, and could have gone on to merge with NGC 1052. An infalling stream of dwarfs being stripped would not face this same constraint since it would have a separate origin from the group without an initial kinematic link to NGC 1052. Still, it is unclear under this picture why so many dark matter free galaxies would be so spatially and kinematically distinct from the group despite being already tidally stripped. Thus, the new results of this work do not fully exclude the exotic forms of alternative models that can reproduce the trail. Even so, as discussed above, the bright, the monochromatic globular clusters of DF2 and DF4 have been reproduced only in collision simulations and, as it stands, formation via a bullet dwarf collision is the only proposed mechanism which explains all properties of DF2, DF4, and the associated trail of galaxies.

We note that these collisions are expected to be somewhat common. \citet{2020ApJ...899...25S} identified 248 collisions from among 95 halo catalogs in TNG100-1 which would produce dark matter free galaxies. \citet{2024ApJ...966...72L} found that 10 of these would result in a trail of dark matter free galaxies directly analogous to the NGC 1052 system, matching the approximate age and also the distance from the host. This contrasts with the Bullet Cluster, which is an extremely rare event in standard cosmological models (\citealt{2006MNRAS.370L..38H,2007MNRAS.380..911S}). Observationally, at least one other galaxy-scale collision in which a similar linear trail of gas was separated from galaxies due to a high speed collision has been identified \citep{2008ApJ...687L..69K}, although in that case no new stars were formed out of the gas.

The question still remains under the bullet dwarf scenario which galaxies constituted the progenitor `bullets.' The findings of \citet{2024ApJ...966...72L} suggest that one of the two progenitors may be becoming, or has already become, a satellite of NGC 1052, while the other may be near one end of the trail traveling at the highest velocity of the system. This would imply one progenitor is simply a `normal' quenched satellite of NGC 1052, making identification somewhat difficult. If it is still aligned with the trail, e.g. due to conservation of angular momentum, DF7 and LEDA 4014647 are possibilities. RCP32 is a good candidate for the high velocity progenitor, as already suggested by \citet{2022Natur.605..435V}. While we find that its radial velocity is 22 km s$^{-1}$ below that of DF2, it is important to remember that we have not measured the velocity of the galaxy itself, but one of its globular clusters moving in the potential of RCP32 with its own peculiar velocity. If RCP32 is indeed dark matter rich, and has a halo which would have normally been associated with half the stellar mass of the entire trail (excluding DF7 and LEDA 4014647), its expected dispersion would be $\approx$40 km s$^{-1}$. Thus, the observed globular cluster radial velocity could be consistent with RCP32 being a dark matter rich progenitor moving at a higher radial velocity than DF2.


\subsection{Globular Cluster Associations} \label{Sec:GCs}

The association of DF9 and LEDA 4014647's clusters is clear based on their kinematic match to the radial velocity for the diffuse light of the galaxies, as well as their morphological match given the brightest clusters' location at the center of the galaxies. This is slightly less clear in the case of RCP32 given that we do not have other radial velocities for comparison since the diffuse light of the galaxy was too faint for observation and the second globular cluster candidate in the slit turned out to be a star. We therefore consider the case that this single RCP32 globular cluster is a chance interloper from NGC 1052. While the large mass and rich globular cluster system of NGC 1052 make it possible, if less likely, for a cluster to have such a velocity (indeed, 4 of 77 globular clusters with known radial velocities are near or above that of RCP32's; \citealt{2019MNRAS.489.3665F}), the distance of RCP32 to NGC 1052 on the sky is ${\approx}22\arcmin$, i.e. $1320\arcsec$, which is $\approx$60 times greater than NGC 1052's effective radius of 21.9$\arcsec$. To compare, the most distant known Milky Way globular cluster is 24 effective radii away \citep{2016ApJ...822...32W}, and while the globular cluster systems of early-type galaxies can be 10 times more extensive than their host galaxy \citep{2019A&A...625A..32B}, this still means that the globular cluster would be a geometric outlier at 6 effective-globular cluster radii away. In addition, RCP32 has a radius of 23$\arcsec$, occupying only $r/4R \approx 0.4{\%}$ of the area of the circular shell at $R \pm r=1320\arcsec{\pm}23{\arcsec}$ from NGC1052. Taken together, the odds of a 2$\sigma$ velocity outlier from NGC 1052 at such an extreme spatial offset being coincident with the dwarf galaxy, which itself takes up a small patch of sky, is effectively nil.


\subsection{Low Surface Brightness Radial Velocities} \label{Sec:RCP32}

As noted by \citet{2021A&A...656A..44R}, RCP32 stands out as an extremely low surface brightness galaxy, similar to BST1047+1156 \citep{2024ApJ...964...67M} and objects from \citet{2020ApJ...899...69L}, if still far brighter that objects in the Local Group \citep{2019MNRAS.488.2743T,2024ApJ...961..126M}. We are pleased to have obtained a redshift for this source. Though in this work we employed novel methodologies to obtain radial velocities for diffuse objects, ultimately it was a single globular cluster in RCP32 that led to its radial velocity determination. Such alternatives to direct redshifts from diffuse light, which require long integration times, remain powerful tools (as in HI in the case BST1047+1156), and similar approaches may be taken for cluster rich samples among \citet{2020ApJ...899...69L}. 

Nevertheless, our novel Ca II H and K line observation, using a high resolution grating and binning heavily to measure radial velocities in reasonable integration times, stands out as an excellent method, allowing us to obtain redshifts for similarly diffuse objects without globular clusters. In particular, we highlight the redshift of DF5, a faint puff as seen in Figure~\ref{Fig:DF45}, for which we obtained a clear Ca II H and K signal. We are continuing to employ this method on diffuse objects.


\subsection{Comparison to Previous Works} \label{Sec:Comparison}

We note that two of our targets, DF9 and LEDA 4014647, already had radial velocity measurements. DF9 had a measurement of 1680 km s$^{-1}$ \citep{2023MNRAS.524.2624G}, $\approx$30 km s$^{-1}$ greater than our LRIS narrow slit mode result of 1648$^{+1}_{-1}$ km s$^{-1}$ for the same galaxy. However, in correspondence the author notes that the uncertainty stated in the abstract of 10 km s$^{-1}$ does not include the $\approx$40 km s$^{-1}$ uncertainty in wavelength calibration, which they suggest we add in quadrature. This adjusted result of 1680$\pm$40 km s$^{-1}$ is indeed in agreement with our new result. LEDA 4014647 had a measurement of 1665$\pm$54 km s$^{-1}$ from an early data release of SDSS, for which our result of $1596^{+3}_{-2}$ km s$^{-1}$ is only in mild, 1.2$\sigma$ tension, and would equally place LEDA 4014647 as an interloper.

\citet{2025ApJ...978...21T} found that the trail galaxies have similar, older ages compared to dwarfs in the group. Indeed, all the galaxies we find to follow the expected radial velocity trend including DF5, Ta21-12000, DF2, RCP26, DF9, and DF4 have ages between 7 and 11 Gyr, while DF7 and LEDA 4014647, the two interlopers identified in this work, do not \citep{2025ApJ...978...21T}. That is, the galaxies we confirm to be on the trail based on their radial velocities have mass-weighted age estimates between consistent with previous 9$\pm$2 Gyr age estimates for DF2 \citep{2018ApJ...856L..30V,2019A&A...625A..77F}, whereas the galaxies we identify as interlopers, DF7 and LEDA 4014647, have ages of $<$7 Gyr and $>$11 Gyr, respectively.\footnote{Though we note the large uncertainties on the DF7 age result do not exclude it on this basis alone. Similarly, RCP21 and RCP28, two of the galaxies not targeted in this work being far too faint for reliable radial velocity determination, technically had age estimates just under 7 Gyr as well, with wide uncertainties of 3-4 Gyr which exclude decisive analysis. Likewise, RCP17 and RCP32 were too faint for any meaningful age constraints.} We also note that \citet{2025ApJ...978...21T} found that the position angles of the trail galaxies are aligned, as determined from the central regions in \textit{HST} imaging, and that this alignment that extends to the exteriors of DF2 and DF4 as seen in deep optical imaging \citep{2022ApJ...935..160K}.\footnote{This may be significant as the dynamical time is especially long in this region (several billion years) due to low density of these spatially extended galaxies lacking dark matter, meaning that the early history of the galaxies may have left a residual impact visible today. It is theoretically possible similar conditions at early stages in DF2 and DF4's formation are responsible for their morphologically similar exteriors if the galaxies indeed shared a common origin, for instance if they experienced a similar external gravitational field (though this point requires closer analysis).}









\section{Summary and Conclusion} \label{Sec:Summary and Conclusion}

In this work we measured radial velocities for seven galaxies on a linear trail with DF2 and DF4, the remarkable galaxies lacking dark matter \citep{2018Natur.555..629V,2019ApJ...874L...5V}, which is theorized to have formed as the result of a high-speed bullet dwarf collision (a small scale analog of the Bullet Cluster; \citealt{2006ApJ...648L.109C,2022Natur.605..435V}). We employed a variety of spectroscopic techniques, obtaining measurements for galaxies with effective surface brightnesses up to 28.6 mag arcsec$^{-2}$. We found that five of these seven galaxies follow the precise kinematic trend predicted by the bullet dwarf collision scenario, to a high degree of significance (with a 2\% chance of random occurrence). Their velocities are highly distinct from the NGC 1052 group, marking the trail as a unique, kinematically connected system. Moreover, the trail's velocity trend is a specific prediction of collisional formation based on the velocities of DF2 and DF4, providing strong support for the bullet dwarf theory. 

High velocity collisions represent an exciting, novel channel of galaxy formation, producing galaxies from the top down in a shock-compressed environment without dark matter. Indeed, bullet dwarfs provide a useful check for the results of such extreme star formation modes in galaxy formation models. Moreover, bullet dwarfs probe the fundamental nature of dark matter, challenging modified gravity on the same galaxy scales where MOND is most relevant and potentially offering an avenue to constrain the cross section of dark matter self-interactions.


\section*{Acknowledgments} \label{Sec:Acknowledgments}
Support from STScI grants HST-GO-14644, HST-GO-15695, and HST-GO-15851, and HST-GO-16912 is gratefully acknowledged. A. J. Romanowsky was supported by National Science Foundation grant AST-2308390. Some of the data presented herein were obtained at Keck Observatory, which is a private 501(c)3 non-profit organization operated as a scientific partnership among the California Institute of Technology, the University of California, and the National Aeronautics and Space Administration. The Observatory was made possible by the generous financial support of the W. M. Keck Foundation. This research is based on observations made with the NASA/ESA Hubble Space Telescope obtained from the Space Telescope Science Institute, which is operated by the Association of Universities for Research in Astronomy, Inc., under NASA contract NAS 5–26555. These observations are associated with program 14644, 15695, 15851, and 16912. M. A. Keim thanks E. Shin and J. Lee for their work and helpful conversations - indeed, in this case, theorists played an important role in predicting a galaxy trail that turned out to exist in deep imaging - and J. Gannon for correspondence on the uncertainty in DF9's radial velocity.

\vspace{5mm}

\textit{HST} data presented in this paper can be accessed from the Mikulski Archive for Space Telescopes (MAST) at the Space Telescope Science Institute via \dataset[10.17909/hj6b-7268]{https://doi.org/10.17909/hj6b-7268}.

\facilities{Keck (LRIS), Keck (KCWI), and \textit{HST} (ACS).}

\software{{\tt PypeIt} \citep{2020JOSS....5.2308P}, KCWI DRP, and astropy \citep{2013A&A...558A..33A,2018AJ....156..123A}.}


\appendix


\section{A Toy Model Simulation of the Expected Trend} \label{Sec:ToyModel}

In order to more fully probe the expected velocity trend along the trail, here we provide a simple kinematic simulation of test particles expanding out after being produced in a bullet collision. As in the main text, the central expectation from a collision scenario is that all trail galaxies originated from approximately the same location and that, due to variations in their post-collision accelerations and velocities, they spread out over time resulting in their current positions on the trail today. Therefore, galaxies on the trail ought to follow a rough kinematic trend corresponding to their spatial position on the line, anchored by the positions and velocities of DF2 and DF4.

It is unlikely the galaxies would remain perfectly aligned in velocity space after several billion years (due to the physics of the collision itself and random motions as well as orbital evolution and gravitational influence of the massive galaxy NGC 1052; see \citealt{2024ApJ...966...72L}, with further discussion below). It is therefore helpful to get some picture of what scatter we might expect in this trend. Given that the positions of galaxies on the trail are very well aligned, the spread in velocity space cannot be too great or else this structure could not have been maintained. Indeed, there is an intrinsic link between the tightness of the spatial distribution of trail galaxies and the expected spread in velocity space we might explore.

\citet{2024ApJ...966...72L}'s excellent work provides a highly useful guide as reviewed in the main text, and indeed we see the scatter from the trend in their TM2 simulation is highly reminiscent of our own findings. Their TM1 and TM3 simulations with higher scatter driven by a high velocity outlier may also be informative, suggesting that the apparent kinematic interlopers DF7 and LEDA 4014647 could be genuine trail members (which can be best determined by their dark matter content). Even so, a simple toy model can provide additional utility. First, by using a much larger set of test particles, we may more fully characterize the potential distribution of velocities and positions of trail galaxies. Second, and most importantly, we may compare to the observer scatter of galaxy positions on the sky, to help better assess whether DF7 and LEDA 4014647 are indeed interlopers.

\begin{figure}
    \centering
    \includegraphics[width=0.47\textwidth]{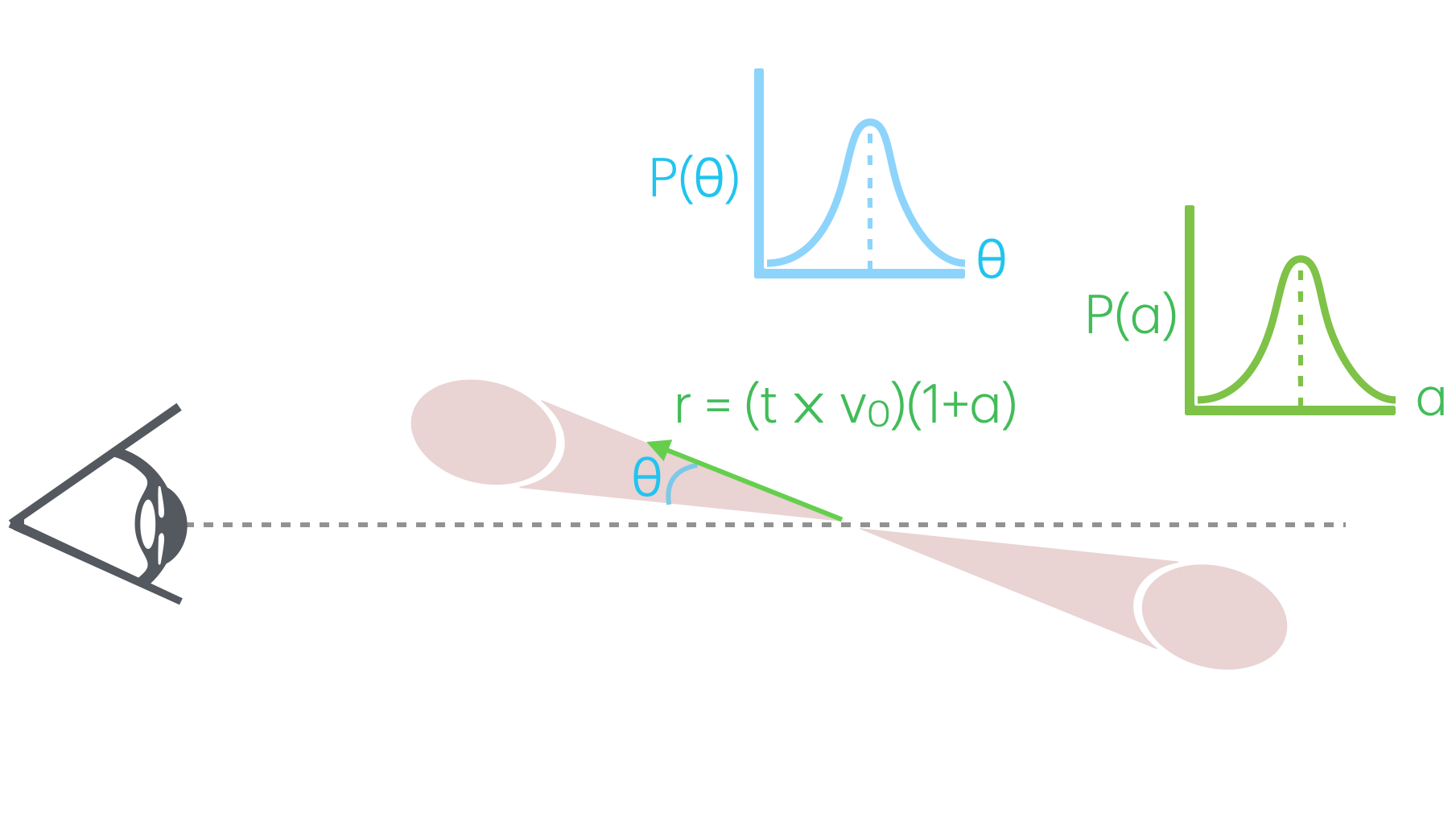}
    \caption{An illustration of how particles are launched in our toy model, with velocity vectors assigned based on opening angles drawn from a Gaussian with a characteristic $\hat{\theta}$ and with effective accelerations drawn from a Gaussian with a characteristic $\hat{a}$.\label{Fig:Cartoon}}
\end{figure}

To explore this expected radial velocity trend and how it may relate to the tight spatial alignment of the galaxies on the trail, we conducted a basic simulation of test particles moving from a common origin at a range of velocities and accelerations, then examined their resulting positions and radial velocities projected onto the sky and line-of-sight. The simulation is kinematics only - it does not exclude an explicit external potential, but does include an effective acceleration term drawn from a random distribution. This toy model simulation is constructed as follows:

First, we created 100,000 test particles and assigned them a random velocity $v_{\rm 0}$ along a collision axis. The magnitude of $v_{\rm 0}$ was drawn from a uniform distribution sufficiently large to probe the full distribution of the particles around the observed spatial extent of the trail ($\pm$650 km s$^{-1}$, which we empirically found to be satisfactory to fully characterize the possible trajectory of trail galaxies within the NGC 1052 field after several Gyr). Second, perpendicular to this axis, oriented at a random azimuthal angle $\phi$, we assigned the particles a transverse velocity component $v_{\rm t}$. The magnitude of $v_{\rm t}$ is defined based on an opening angle $\theta$, such that $v_{\rm t} = v_{\rm 0} \tan{\theta}$, where $\theta$ is drawn from a Gaussian distribution. This Gaussian distribution has a mean of $0^\circ$ and a standard deviation of $\hat{\theta}$, a characteristic opening angle. This transverse component causes particles to spread out on the sky and is the principal link between the scatter of galaxies in position and velocity space. Third, we add a random acceleration term to account for the fact that the particles would experience unique accelerations over the course of the simulation causing non-linear motions. These include the combined gravitational pulls of the other trail galaxies and galaxies within and nearby the NGC 1052 group, resulting a unique effect for each galaxy on the trail. We characterize this acceleration in terms of a unit-less factor $a$ of the velocity divided by half of the simulation time $\Delta t$, such that the true acceleration in physical units is $a \times v_0/(\Delta t/2)$. In this way, the normal kinematic equations for displacement and velocity become $\Delta r = v_0 \Delta t \times (1+a)$ and $v_f = v_0 \times (1+2a)$. This unit-less factor $a$ is drawn from a Gaussian distribution. This Gaussian distribution has a mean of $0$\%, i.e. an acceleration of $0$, and a standard deviation of $\hat{a}$, i.e. an acceleration of $\hat{a} \times v_0/(\Delta t/2)$. A visual depiction of the opening angle and acceleration factors is given in Figure~\ref{Fig:Cartoon}.

We compute the final location and velocities of particles as projected from the frame of the collision onto the sky and along the line-of-sight using the following relations:
\begin{equation} \label{Eq:xsky}
x_{\rm sky} = \arctan{ \biggl\{ \left( v_{\rm 0}\sin{(i)}+ v_{\rm t} \cos{(\phi)} \cos{(i)} \right) \biggr. } \biggl. \times \left( \frac{D}{\Delta t (1+a)} + v_{\rm 0} \cos{(i)}+v_{\rm t} \cos{(\phi)} \sin{(i)} \right)^{-1} \biggr\},
\end{equation}
\begin{equation} \label{Eq:ysky}
y_{\rm sky} = \arctan{ \biggl\{  v_{\rm t} \sin{(\phi)} \biggr. } \times \biggl. \left( \frac{D}{\Delta t (1+a)} + v_{\rm 0} \cos{(i)} + v_{\rm t} \cos{(\phi)} \sin{(i)} \right)^{-1} \biggr\} ,
\end{equation}
and
\begin{equation} \label{Eq:vlos}
v_{\rm los} = \left( v_{\rm 0} \cos{(i)} +  v_{\rm t} \cos{(\phi)} \sin{(i)} \right) (1+2a) + v_{\rm los, 0},
\end{equation}
where $x_{\rm sky}$, $y_{\rm sky}$, and $v_{\rm los}$ are the output projected angular sky positions and line-of-sight velocity, $D$ is the distance to the center of the trail, taken to be 21 Mpc \citep{2022Natur.605..435V}, $i$ is the inclination angle relative to the line-of-sight, inferred from the sky projected separation between DF2 and DF4 and their relative distances to be 7$^{\circ}$ \citep{2020ApJ...895L...4D,2023ApJ...957....6S}, and $v_{\rm los, 0}$ is the radial velocity of the collision site at the trail center (this is a few km s$^{-1}$ off from the mean velocity of DF2 and DF4, due to the effects of sky projection).

Eq.~(\ref{Eq:xsky}) takes into account motion on the sky both due to motion along the collision axis from $v_{\rm 0}$ and motion due to the transverse velocity component $v_{\rm t}$ which may be appear on the same axis in projection depending on the random angle $\phi$. Similarly, Eq.~(\ref{Eq:ysky}) accounts for motion in the direction perpendicular to the collision axis due to traverse motion $v_{\rm t}$, depending on the random angle $\phi$. Both Eq.~(\ref{Eq:xsky}) and Eq.~(\ref{Eq:ysky}) account for the change in the distance of particles due to motion of $v_{\rm 0}$ or $v_{\rm t}$ projected onto the line-of-sight. Finally, Eq.~(\ref{Eq:vlos}) takes into account both the velocity along the collision axis $v_{\rm 0}$ projected onto the line-of-sight as well as the contribution of the transverse motion $v_{\rm t}$, depending on the random angle $\phi$.

\begin{figure*}
    \centering
    \includegraphics[width=0.47\textwidth]{ 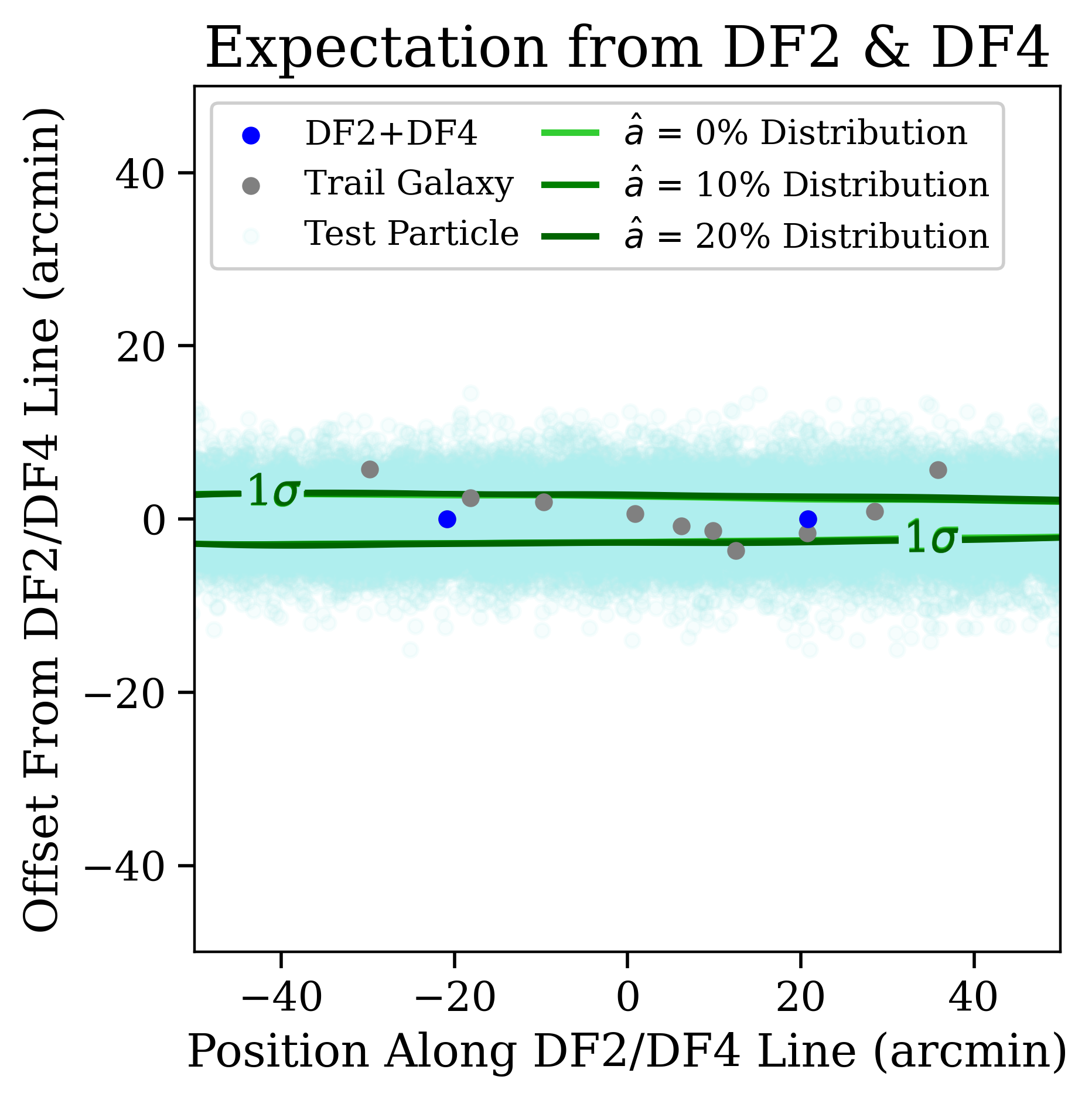}\quad\includegraphics[width=0.480017761989343\textwidth]{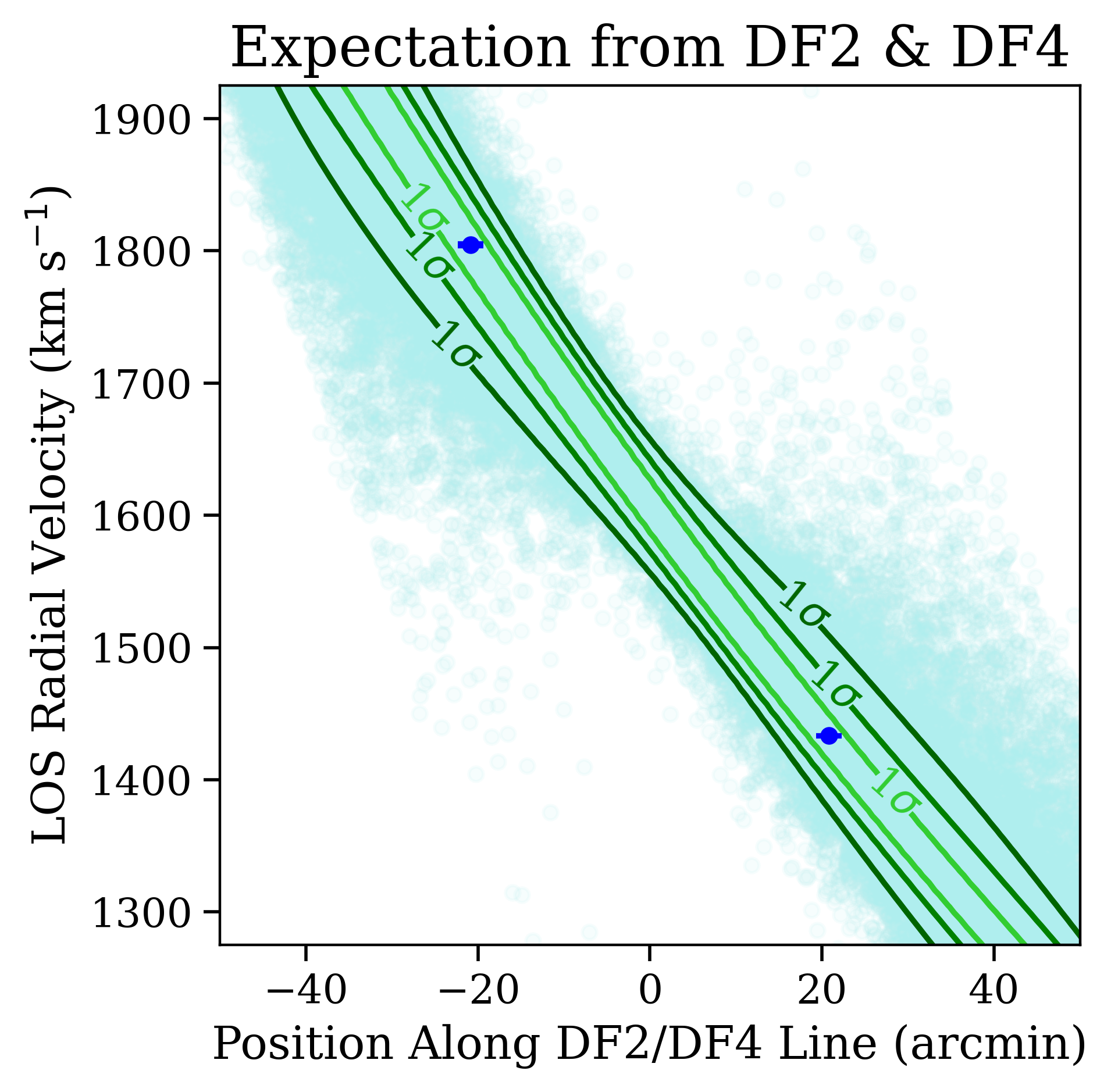} 
    \includegraphics[width=0.47\textwidth]{ 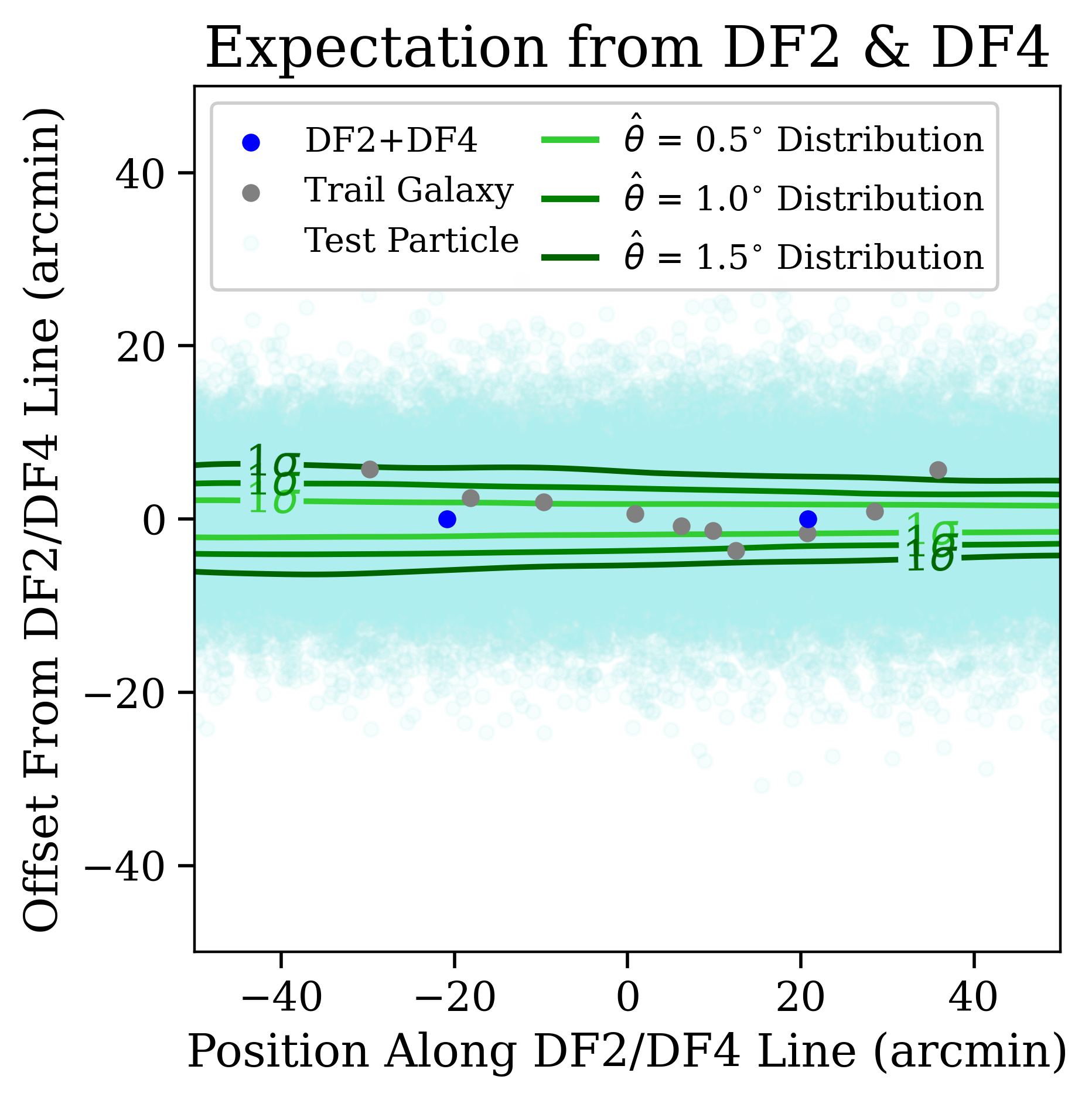}\quad\includegraphics[width=0.480017761989343\textwidth]{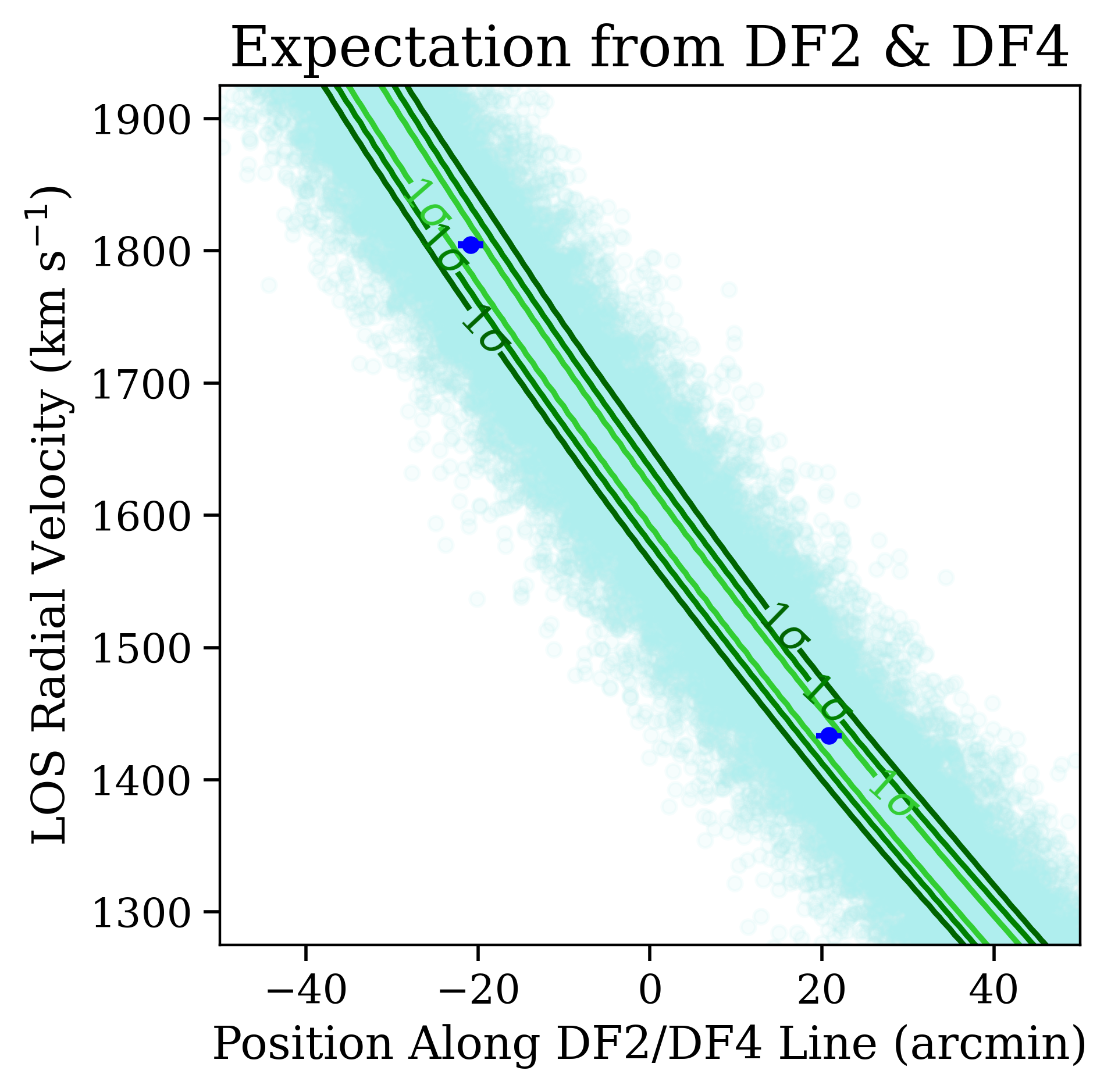} 
    \caption{The distribution of 100,000 particles in a simulated bullet dwarf collision both in position projected on the sky (\textit{left panels}; $x_{\rm sky}$ and $y_{\rm sky}$) and velocity projected along the line-of-sight (\textit{right panels}; $x_{\rm sky}$ and $v_{\rm los}$). Density contours contain 68.27\% of simulated particles, which are themselves given in light turquoise, and are colored according to either the effective acceleration factor $a$ (\textit{top panel}s) or the opening angle $\theta$ (\textit{bottom panels}). DF2 and DF4's positions and radial velocities with associated uncertainties are indicated in blue (note that the error bars are vertical, not horizontal, representing small uncertainties). Positions are given in coordinates relative to the line connecting DF2 and DF4 (above or below this line on the y-axis, and along this line relative to the midpoint between DF2 and DF4 on the x-axis), i.e. the projected collision axis.\label{Fig:Simulation}}
\end{figure*}

$\Delta t$ determines how far a particle with a particular velocity will travel and therefore the relation between position and velocity (with a lower $\Delta t$ corresponding to a steeper kinematic trend). We select $\Delta t$ based on the time at which the radial velocities of test particles at the present day positions of DF2 and DF4 matches their observed radial velocities, 6 Gyr. This is below the observed age of DF2, 9$\pm$2 Gyr (\citealt{2018ApJ...856L..30V}; \citealt{2019A&A...625A..77F}; \citealt{2025ApJ...978...21T}), which presents a mild tension since DF2 should have an age approximately equal to $\Delta t$, given that it would have formed shortly after the collision. However, below we alter the simulation time to match the observed trend for all galaxies on the trail (finding a result in better agreement with the observed ages). The resulting projected distribution of particles at this time is given in Figure~\ref{Fig:Simulation}, with density contours containing 68.27\% of particles for simulations for various values of $\hat{\theta}$ and $\hat{a}$. The actual sky positions of trail galaxies are given in the left panel, and the observed radial velocities of DF2 and DF4 are given in the right panel.

In Figure~\ref{Fig:Simulation}, we explore the two tunable parameters of the model, $\hat{\theta}$ and $\hat{a}$ (the other parameters of the model, as described above, are taken from observation). We see that $\hat{\theta}$ strongly affects the spread of particle positions and velocities, while $\hat{a}$ primarily affects the observed velocities. Indeed, the acceleration term can cause a mis-ordering that does not noticeably affect the vertical width of the spatial distribution but widens the distribution in the velocity spread. This is because the acceleration term, as visualized in Figure~\ref{Fig:Cartoon}, cannot change the direction a particle is heading, but only the time it takes to get there and the velocity it is observed to be traveling at. Since it can have either a positive or negative effect, and since we probe so wide a range of $v_{\rm 0}$, the overall vertical width of the distribution of particles is unaffected by $\hat{a}$. However, for a larger $\hat{a}$, at a particular position there are many more particles with observed velocities that do not match the basic linear expectation, thereby widening the distribution in $v_{\rm los}$.

Thus, under our model $\hat{\theta}$ may be fixed based on the observed spatial distribution of the trail. We find that the distribution of the observed trail is best characterized by $\hat{\theta} = 0.75^\circ$. While this angle is small, the trail is quite long and thin and, moreover, we have defined $\theta$ relative to $v_{\rm 0}$, a parameter which was drawn from a uniform distribution purposefully made to be overly large. This modeling decision was made to better characterize the expected trend at all locations, without the particle distribution falling off at points where particles would not have sufficient $v_{\rm 0}$ to reach. However, as a result the physical implications of our constraint on $\hat{\theta}$ loses some meaning. Still, such an opening angle could not be too large in any case given the trail is so long and tight, with a high geometric significance as characterized by \citet{2022Natur.605..435V} and upcoming work by J. An et al. (in prep) showing that the tight linear nature of the trail relative to its local environment is unique among wider diffuse object catalogs. $\hat{a}$, on the other hand, can only be constrained by a greater sample of trail galaxy radial velocity measurements.

Our model is a simplified, parameterized version of the full hydrodynamical model of \citet{2024ApJ...966...72L}. They similarly conclude that dark matter deficient galaxy trails formed from a high-speed collision should follow a roughly linear trend in radial velocity as a function of position along the trail. They find that the radial velocities of simulated trail galaxies are sometimes offset from the linear expectation due to baryonic physics and orbital collision geometries, which is encapsulated by our random acceleration term. Note that \citet{2024ApJ...966...72L} also find that the dark matter rich remnants of progenitor galaxies typically do not follow this trend, and generally are significantly offset from it. The dark matter free trail galaxies were formed from gas that separated from the two colliding progenitor galaxies through bullet cluster-like shocks and compression. The trail galaxies are therefore expected to have a lower velocity than the remnants of the progenitor galaxies. Specifically, the remnant on the higher velocity side would typically have a higher velocity than the trail galaxies, whereas the remnant on the lower radial velocity side may have fallen into the group or NGC 1052 itself (J. Lee, private communication).

A primary utility of our custom simulation is to connect the observed spatial scatter on the sky to the expected scatter in velocity space. This is especially helpful in outlier determination. While \citet{2024ApJ...966...72L} explore the expected trend in several case, they do not compare to the spatial scatter observed on the sky. It therefore unclear if the galaxies in the TM1 and TM3 simulations of \citet{2024ApJ...966...72L} which have higher scatter from the trend are representative of our trail. Given that the observed trail is so tight on the sky, we can use our toy model here to comment on what degree of scatter is expected.

\begin{figure*}
    \centering
    \includegraphics[width=0.49413413675\textwidth]{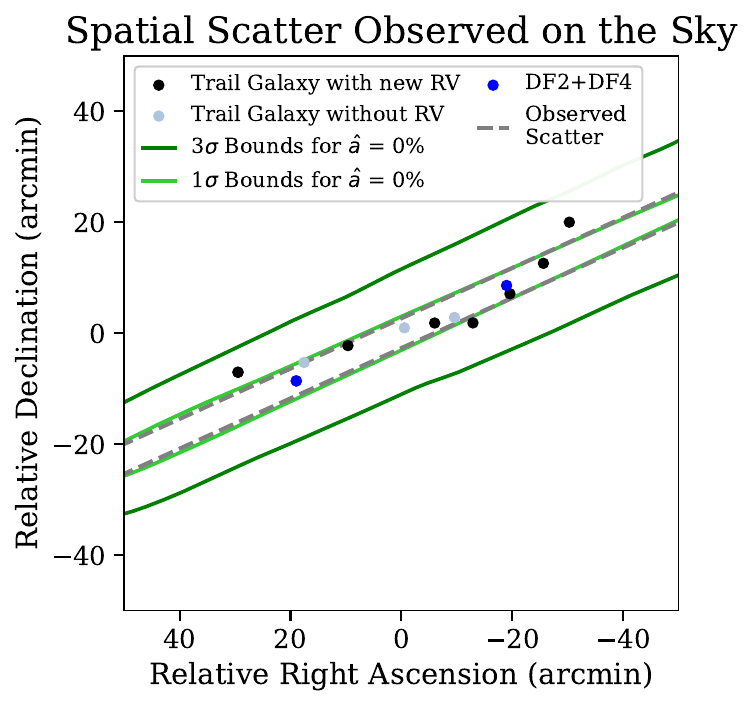}\quad\includegraphics[width=0.47\textwidth]{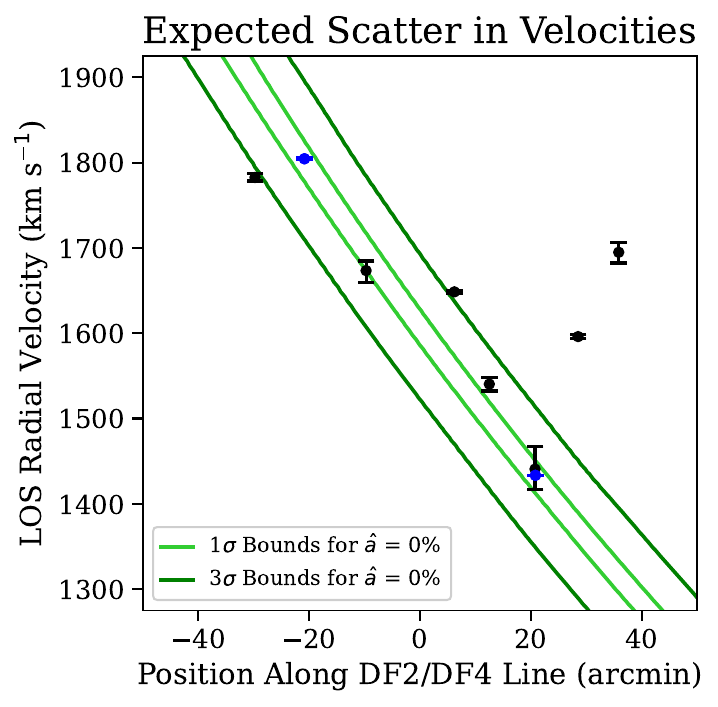} 
    \caption{Outlier determination as based on the constant motion $\hat{a} = 0\%$ model from Figure~\ref{Fig:Simulation}. $\hat{\theta} = 0.75^\circ$ has been tuned such that the 1$\sigma$ bounds of test particles positions (\textit{lime green solid lines}) matches the observed scatter on the sky (\textit{grey dashed lines}). The measured velocities from the main text of the paper (\textit{black circles}) are compared to the 3$\sigma$ bounds of test particles velocities (\textit{dark green solid lines}), indicating the rightmost galaxies are likely to be interlopers.\label{Fig:NoAcc}}
\end{figure*}

With this in mind, in Figure~\ref{Fig:NoAcc} we consider the new velocities measured in this work in the context of the distribution of simulated test particles which best matches the spatial scatter as observed on the sky, in order to comment on which galaxies our one of the 2$\pm$2 interlopers predicted by \citet{2022Natur.605..435V}. To do so, we explore the wide distribution of simulated test particles without random accelerations, i.e. moving with constant speeds, labeling galaxies that extreme outliers as interlopers. Setting $\hat{a} = 0\%$, five of the new measurements fall within the `3$\sigma$' bounds of the simulated distribution of velocities containing 99.73\% of test particles (we increase the number of test particles to 200,000 to better characterize these bounds) and have velocities approximately equal test particles included in the simulation. However, two do not, LEDA 4014647 and DF7, and instead are above their expected velocities by $>$100 and $>$200 km s$^{-1}$ more than any included test particle. They stand as 5$\sigma$ and 9$\sigma$ outliers, respectively. We therefore determine LEDA 4014647 and DF7 are interlopers and not true members of the trail.

As discussed above, this is unrealistic: the galaxies will experience unique accelerations as they travel through a dynamical gravitational field and will not move with constant speed following the collision. As a result, these velocity distribution bounds are overly tight. However, it does give us the basic expectation based on the distribution of trail galaxies on the sky, and a strong indication that LEDA 4014647 and DF7 are interlopers.

Next, we consider our simulation model as applied to our measurements, in incorporating more realistic physics while treating LEDA 4014647 and DF7 as interlopers. In Figure~\ref{Fig:Trend}, we find that $\hat{a} = 17\%$, i.e. accelerations drawn from a Gaussian with a standard deviation of $17/100\times v_0/(\Delta t/2)$, best fits the observed rms relative to the expectation from DF2 and DF4 (i.e. the scatter relative to the expected trend, calculated as the root mean square of the observed velocities -- excluding interlopers -- after subtracting the expected velocities from DF2 and DF4 at each galaxy's position; this rms is added and subtracted to the expected trend and plotted in grey dashed lines). We find $\hat{\theta} = 0.4^\circ$ and $\hat{a} = 13\%$, or an acceleration of $15/100\times v_0/(\Delta t/2)$, is able to match the decreased rms relative to the fit line (i.e. the scatter relative to the fit line, calculated in the same manner as described above). As expected from \citet{2024ApJ...966...72L}, this is greater than what would be expected from constant linear motion. Note that we extend our simulation to end at $\Delta t = 7$ Gyr, in order to match the slightly less steep trend when considering all the observed radial velocities in this larger sample together. This is in better agreement with the ages of the galaxies (\citealt{2018ApJ...856L..30V}; \citealt{2019A&A...625A..77F}; \citealt{2025ApJ...978...21T}), though a simulation with more complex geometries, e.g. \citet{2024ApJ...966...72L}, would be necessary for more exact age comparisons.

As a final consistency check on our outlier determination, we note that of the 100,000 test particles in the toy model simulations shown in Figures~\ref{Fig:Trend} and \ref{Fig:TrendAdjusted}, almost none appear near DF7 and only a handful lie near LEDA 4014647, whereas the other galaxies are in regions densely populated by test particles. Given that we have a sample of 12, it is unlikely (though not wholly impossible) that these are genuine trail members.








\bibliographystyle{aasjournal}
\bibliography{bibliography}

\end{document}